\documentclass[12pt]{article}
\usepackage{epsfig}
\usepackage{graphicx}
\usepackage{amsmath}
\usepackage{amsthm}
\usepackage{amssymb}
\usepackage{latexsym}
\usepackage{color}
\usepackage{mathrsfs}
\usepackage{natbib}
\usepackage{hyperref}
\usepackage{appendix}
\usepackage{dsfont}
\usepackage[left=3cm,top=4cm,right=3cm,bottom=3cm]{geometry}
\usepackage{lscape}
\usepackage{rotating}

\usepackage{mathabx}

\theoremstyle{plain}

\newtheorem{theorem}{Theorem}[section]

\theoremstyle{definition}

\usepackage{dsfont}

\usepackage{subcaption}

\DeclareMathOperator{\vop}{vec}

\newcommand{\e}{\mathds{E}}
\newcommand{\R}{\mathds{R}}

\newcommand{\cov}{\mathds{C}\text{ov}}

\newcommand{\xvec}{\boldsymbol}
\newcommand{\xmat}{\mathbf}
\newcommand{\xset}{\mathds}

\DeclareMathOperator{\cls}{cls}

\DeclareMathOperator{\clr}{clr}
\DeclareMathOperator{\ilr}{ilr}

\DeclareMathOperator{\ailr}{i\alpha t}
\DeclareMathOperator{\aclr}{c\alpha t}

\DeclareMathOperator{\si}{\mathds{S}}
\DeclareMathOperator{\E}{E}
\DeclareMathOperator{\Var}{Var}
\DeclareMathOperator{\Cov}{Cov}
\DeclareMathOperator{\tr}{tr}

\newtheorem{assumption}{Assumption}

\graphicspath{{./figs/}}

\begin{document}
\title{\LARGE \bf 
Spatiotemporal Autoregressive Models for Areal Compositional Data}

\date{\date{}}
\author{Matthias Eckardt and Philipp Otto}
\maketitle
\begin{center}
{{\bf Matthias Eckardt$^{a}$} and  {\bf Philipp Otto}$^{b}$}\\
\noindent $^{\text{a}}$ Chair of Statistics, Humboldt Universit\"{a}t zu Berlin, Berlin, Germany\\
\noindent $^{\text{b}}$ School of Mathematics and Statistics, University of Glasgow, United Kingdom, corresponding author: philipp.otto@glasgow.ac.uk
\end{center}

\begin{abstract}
Compositional data, such as regional shares of economic sectors or property transactions, are central to understanding structural change in economic systems across space and time. This paper introduces a spatiotemporal multivariate autoregressive model tailored for panel data with composition-valued responses at each areal unit and time point. The proposed framework enables the joint modelling of temporal dynamics and spatial dependence under compositional constraints, and is estimated via a quasi maximum likelihood approach. We build on recent theoretical advances to establish identifiability and asymptotic properties of the estimator when both the number of regions and time points grow. The utility and flexibility of the model are demonstrated through two applications: analysing property transaction compositions in an intra-city housing market (Berlin), and regional sectoral compositions in Spain’s economy. These case studies highlight how the proposed framework captures key features of spatiotemporal economic processes that are often missed by conventional methods.
\end{abstract}
{\it Keywords: Spatiotemporal compositional data, economic compositions, spatial econometrics, real estate} 

\section{Introduction}

Regional economic systems frequently undergo transformations driven by variations in the contributions of fundamental elements, such as the types of housing or industrial sectors prevalent within them. Formally defined, these fundamental components denote interrelated proportions that collectively form a complete entity, which must total a constant within each regional context. This includes situations like the allocation of economic activities among various sectors or the categorisation of property transactions by their types, and acknowledging this intrinsic dependence structure is crucial when analysing such datasets. Compositional data analysis, as outlined by \citep{MR1873662}, offers a comprehensive methodological framework that facilitates the examination of such constrained relational data. Nonetheless, the analysis of intricate regional economic systems, where these fundamental components significantly influence, remains challenging. Aside from utilising traditional compositional data analysis techniques \citep{doi:10.1002/9781119976462} for the non-spatial components of the data, it is valuable to examine the spatial distribution characteristics of these compositions. A principal assumption in this examination is the influence of geographical proximity on local data, consistent with Tobler's First Law of Geography \citep{tobler}\footnote{To quote Sir Ronald A. Fisher: ``patches in close proximity are commonly more alike, as judged by the yield of crops, than those which are further apart'' \cite{fisher1935design}.}. In addition to the spatial dependence emanating from neighbouring areas, compositional data are frequently recorded multiple times over a period. In such environments, it is necessary to analyse both spatial interactions amidst the compositions and their temporal autocorrelation alongside dynamic progression. However, even with the increased accessibility of complex non-scalar spatio-temporal data and a pronounced need for spatio-temporal areal regression models, there exists a noticeable absence of appropriate model specifications for composition-valued information distributed across regions, i.e., areal spatial data. Although some literature documents applications of Whittle's simultaneous autoregressive \citep{whittle} and Besag’s or Mardia's (multivariate) conditional autoregressive models \citep{Besag:74, mardia} in spatial compositional datasets, extensions to more elaborate data structures are rare. This paper seeks to bridge this gap by presenting a multivariate simultaneous autoregressive model where the response at each site and temporal instance is a composition-valued data point. Methodological advancements, driven by regional economic dynamics data, are exemplified through case studies on the spatiotemporal composition of the Berlin real estate market and diverse economic sectors in Spain \textcolor{black}{(presented in the Appendix)}.

Concentrating on compositions of benthic species with exclusively non-zero parts, \cite{Billheimer1997} implemented an additive log ratio transformation (alr) \citep{10.5555/17272} to position the data within Euclidean space and utilised a multivariate spatial conditional autoregressive model. \cite{Leininger2013} proposed a multivariate conditional autoregressive model to account for spatial configurations in land use and cover amidst multiple zero components. Beyond formulating the aforementioned conditional model, literature also addresses hierarchical \citep{PIRZAMANBEIN201814}, simultaneous autoregressive \citep{Nguyen03042021}, and spatial lag quantile regression models \citep{Zhao11122024}. Furthermore, \cite{Yoshida2018} and \cite{Laurent2023} explored spatial relationships and covariate effects within compositional spatial area models, correspondingly. \textcolor{black}{Turning to compositional data dynamics, \cite{Paciorek2009} develops a Bayesian hierarchical spatio-temporal model to reconstruct historical forest composition from fossil pollen, \cite{Mastrantonio2019} models smooth temporal changes in compositions via a logistic–Gaussian process, and \cite{Kettunen2024} combines latent Gaussian processes with a Dirichlet–multinomial observation model to capture competition for space.}

Yet, to our best knowledge, no models exist to elucidate spatio-temporal dynamics in compositional areal data. Motivated by urban economic applications, this paper introduces a multivariate simultaneous autoregressive model tailored for spatial areal data, where the focus for each spatial site is a sequence of composition-valued quantities tracked over consecutive time steps. The proposed modelling framework, targeting composition-valued areal data observed over time, merges a multivariate simultaneous autoregressive structure, \textcolor{black}{building upon recent multivariate spatiotemporal autoregressive heteroscedasticity models by \cite{otto2024multivariate} and the multivariate simultaneous autoregressive model by \cite{Kelejian1998},} with isometric log-ratio transformations to address compositional constraints. 

The methodology is motivated by constrained regional economic dynamics, providing detailed case studies on Berlin's housing market compositions and sectoral compositions within Spanish municipalities (Section~\ref{sec:motivation}). Estimation utilises quasi-maximum likelihood, with proof of consistency and asymptotic normality as spatial and temporal domains expand (Section~\ref{sec:theory}). A simulation study assesses the estimator's finite-sample characteristics (Section~\ref{sec:sim}), and empirical findings (Section~\ref{sec:appl}) illustrate the model’s capability to represent interpretable spatiotemporal trends in real-world constrained multivariate systems. Discussion on the findings and potential advancements concludes the paper (Section~\ref{sec:conclusion}).

\section{Cases Studies of Constrained Regional Economic Dynamics}\label{sec:motivation}

\begin{figure}
    \centering
    \includegraphics[width=0.32\linewidth]{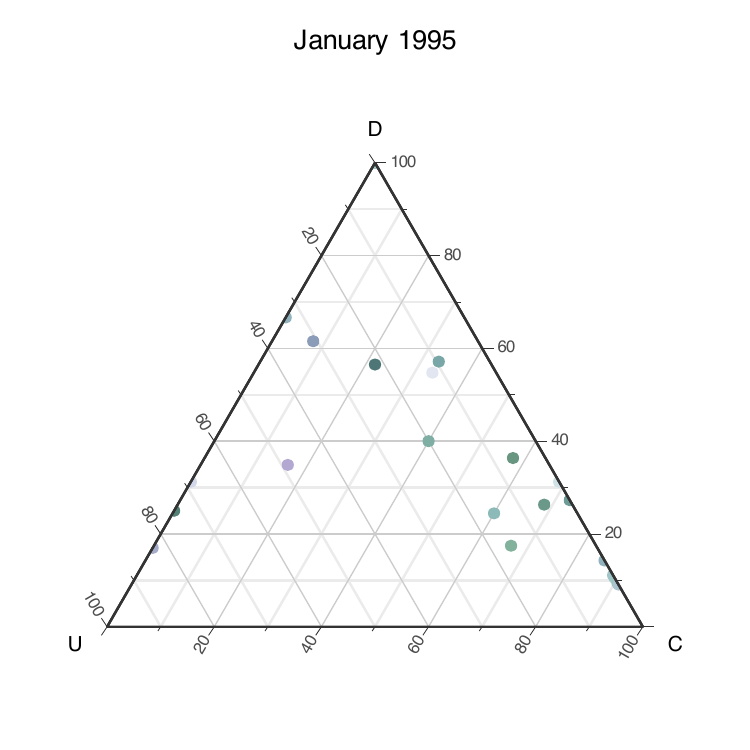}
    \includegraphics[width=0.32\linewidth]{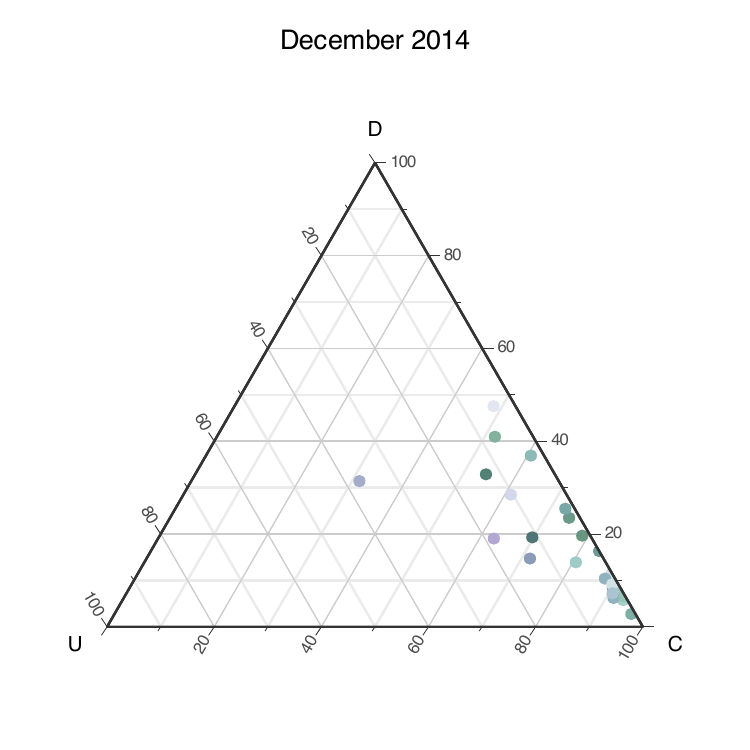} 
    \includegraphics[width=0.32\linewidth]{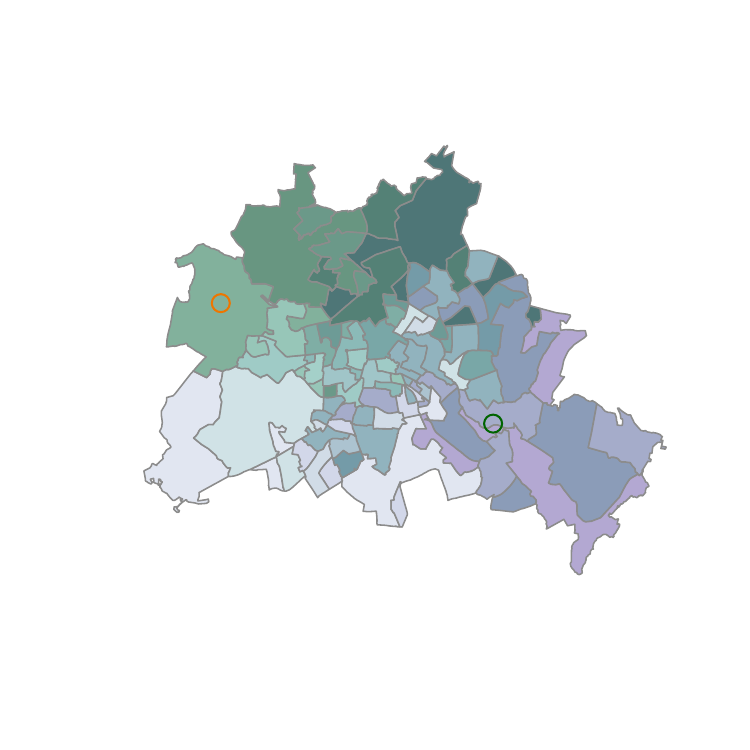} \\
    \includegraphics[width=0.43\linewidth]{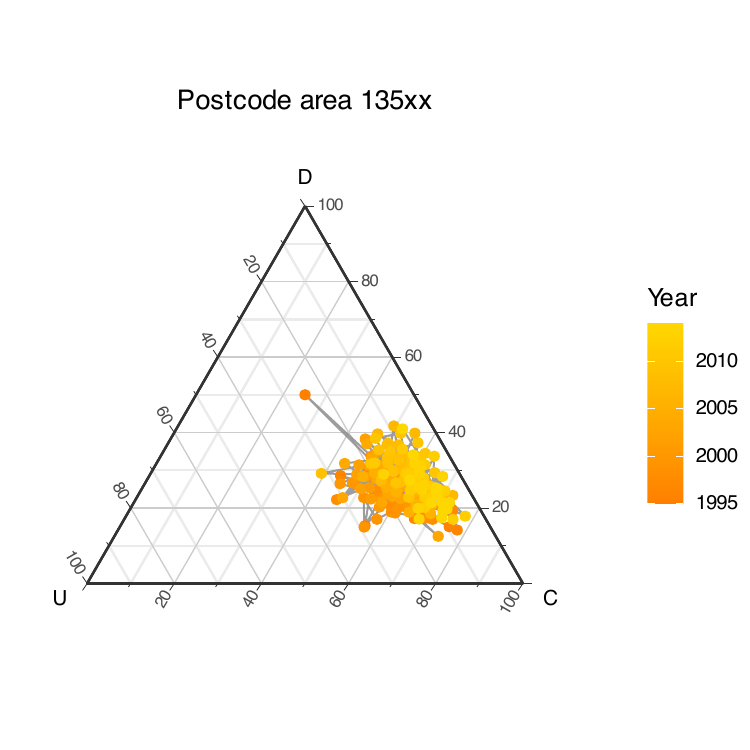}
    \includegraphics[width=0.43\linewidth]{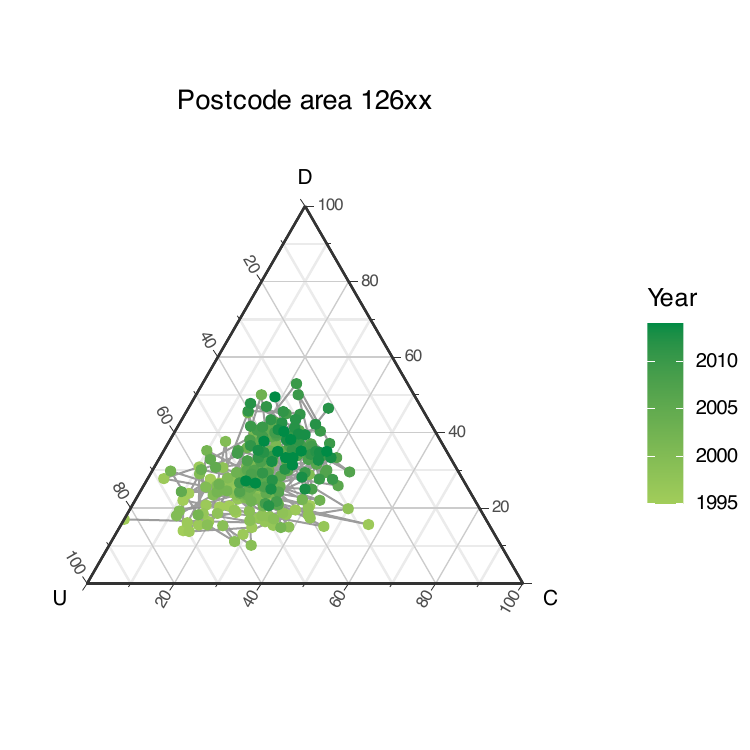}\\
    \includegraphics[width=0.43\linewidth]{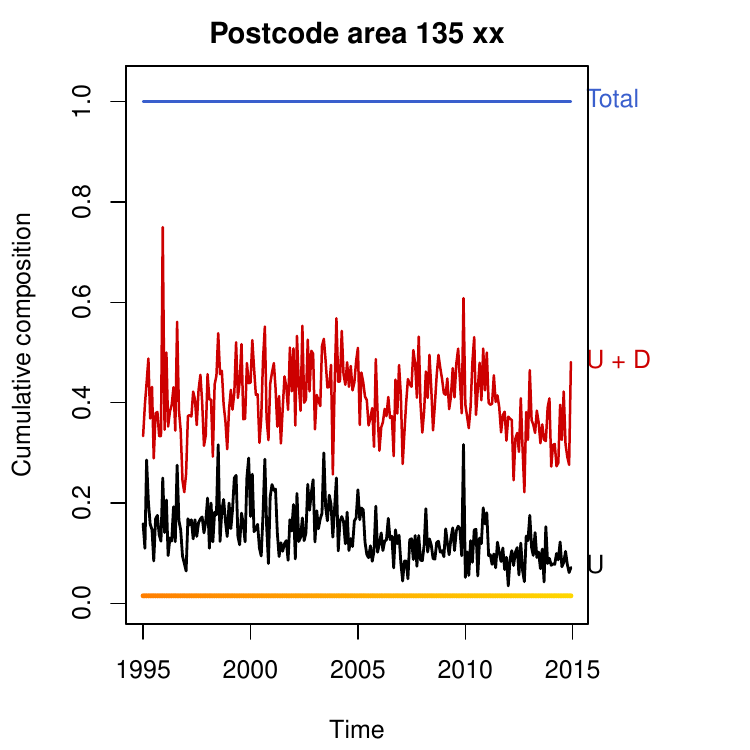}
    \includegraphics[width=0.43\linewidth]{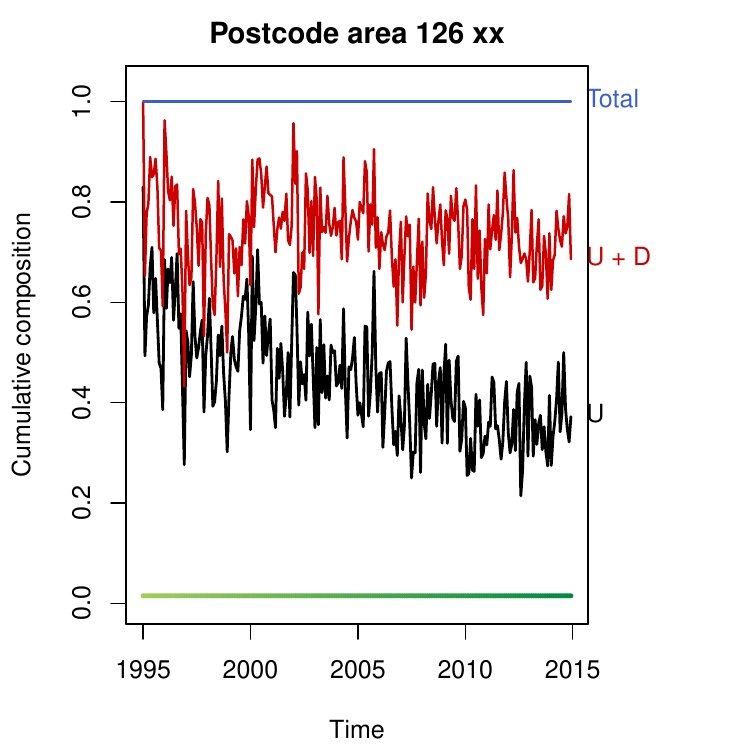}
    \caption{Overview of the Berlin real-estate market composition data set. First row: Compositions of real estate transactions (U: Undeveloped land, D: Developed land, C: Condominium) in January 1995 (left) and December 2014 (centre). The points are coloured according to the spatial location as shown in the map legend (right). Middle row: Composition of two selected regions, where the points are coloured according to the time points. The colours correspond to the colour of the marks on the map. Bottom row: \textcolor{black}{Temporal dynamics of closed 3-part composition for two selected locations over time. Area under black line corresponds to the first, area between black and red to the second, and area above the red line to the third compositional part}.}
    \label{fig:berlin1}
\end{figure}

This research is based on an in-depth analysis of two substantial case studies, which serve to illustrate the complexities and potentials of modelling regional economic dynamics under constraints. The first case study scrutinises the monthly composition of real estate transactions within Berlin's 24 different postcode regions over a span from 1995 to 2015. This analysis concentrates on the proportions of transactions differing among condominiums, developed parcels of land, and undeveloped land within the urban real estate market. This dataset provides invaluable insights into the temporal evolution of distinct sectors within the property market confined to a single city. Furthermore, fluctuations in one market segment, such as a rise in condominium sales, are assessed for their potential impact on the demand dynamics for developed or undeveloped properties in nearby areas. The foremost advantage of this study is the capacity to pinpoint spatial dependencies, indicating how property trends in one postal region might have implications for neighbouring ones, thereby highlighting the intrinsic connectedness of the urban real estate market. Figure~\ref{fig:berlin1} presents comprehensive descriptive overviews of this data set. Analysis of the Berlin data reveals a gradual transformation in the intra-urban housing market composition, alongside indications of moderate spatial autocorrelation among proximate postcode regions. 

The second case study delves into annual data collected from 2012 through 2021, focusing on the allocation of local enterprises across three primary economic sectors within 2,793 municipalities in Spain. These sectors include {services} (encompassing communications, financial and insurance services, administrative and support services, as well as educational, health and social services, and arts, recreation, and entertainment activities), {industry} (comprising extractive processes, manufacturing industries, energy and water provision, sanitation, and waste management operations), and {construction}. \textcolor{black}{The detailed results of the second case study can be found in the Appendix.} The two settings examined herein represent disparate spatiotemporal scales. The data from Berlin offers a high level of temporal granularity across a moderately large geographical area, whereas the dataset pertaining to Spain is characterised by extensive cross-sectional breadth, albeit with a relatively short temporal span. 

These empirical patterns motivate the need for a flexible and interpretable model that jointly captures (i) the constrained nature of the data, (ii) temporal dependence within regions, (iii) spatial dependence across regions, and (iv) cross-component interactions between parts of the composition. In the next section, we introduce a spatiotemporal multivariate autoregressive model that meets these requirements and provides a coherent inferential framework for compositional areal panel data.



\section{Modelling Framework for Spatiotemporal Compositional Data}\label{sec:theory}

This section introduces the modelling framework for spatiotemporal areal data with compositional responses. We begin by formalising the structure of composition-valued observations on spatial units over time (Section~\ref{sec:prelim}), and subsequently present a multivariate spatiotemporal autoregressive model that accounts for both spatial and temporal dependence while respecting compositional constraints (Section~\ref{sec:methods}).

Our approach builds upon recent advances in spatial econometric modelling, in particular the multivariate simultaneous autoregressive (MSAR) model of \citet{yang2017identification}, \textcolor{black}{which itself extends the multivariate simultaneous autoregressive model by \cite{Kelejian1998}}. In that model, the outcome is a matrix of multivariate observations over areal units, and spatial dependence is captured through both within- and cross-component spatial autoregression:
\begin{equation} \label{eq:initialbackground}
 	\xmat{Y}  =  \xmat{W}  \, \xmat{Y} \,  \xmat{\Psi}  + \xmat{X} \,  \xmat{\Pi} + \xmat{V}\, ,
\end{equation}
where \(\xmat{Y}\) is an \(n \times p\) matrix of outcomes for \(n\) spatial units and \(p\) response components, \(\xmat{W}\) is a known spatial weight matrix, \(\xmat{\Psi}\) is a \(p \times p\) matrix of spatial autoregressive coefficients, and \(\xmat{\Pi}\) captures regression effects. The matrix \(\xmat{V}\) contains spatially unstructured noise.

We extend this framework to incorporate temporal dependence, analogous to classical spatial dynamic panel data models \citep{Yu08}. In the univariate case, a widely used model is given by
\begin{equation} \label{eq:yu_model}
 	\xvec{Y}_t  =  \lambda \xmat{W}  \, \xvec{Y}_t +  \gamma \xvec{Y}_{t-1} +  \rho \xmat{W} \xvec{Y}_{t-1} + \xmat{X} \,  \xvec{\beta} + \xvec{c} + \xvec{V}_t\, ,
\end{equation}
where \(\xvec{Y}_t\) is the vector of observations at time \(t\), and the parameters \(\lambda\), \(\gamma\), and \(\rho\) control spatial contemporaneous, temporal, and spatiotemporal lag effects, respectively. Moreover, the series \(\{\xvec{V}_t: t = 1,\ldots, T\}\) contains spatially and temporally independent error vectors.

Our goal is to extend these ideas to the case where \(\xvec{Y}_t\) consists of composition-valued multivariate responses. This requires a formulation that captures multivariate spatial and temporal dependencies, while ensuring that model outputs respect the simplex geometry inherent to compositions. To this end, we introduce a spatiotemporal MSAR model that operates on isometrically log-ratio-transformed compositions, and derive an associated quasi-maximum likelihood estimator suitable for high-dimensional panels.

\subsection{Compositional areal data in space-time}\label{sec:prelim}

Extending compositional data to spatiotemporal areal processes let $\xmat{Z}_t=\lbrace \xvec{Z}_{t}(\xvec{s}_n) \rbrace$ denote a collection of $nT$ constrained vectors which quantitatively describe the relative contribution of $D$ parts of some whole on $\mathcal{S}\times \mathcal{T}$ with $\mathcal{S} = \lbrace\xvec{s}_1, \ldots, \xvec{s}_n\rbrace$ denoting a countable set of $n$ (at least partially) interconnected areal units and $\mathcal{T} =\lbrace t_j\rbrace^T_{j=1}$ a set of $T$ distinct equidistant steps in time. In general, we assume that the areal units in $\mathcal{S}$ and the compositional $D$ parts remain consistent across all temporal instances $t\in \mathcal{T}\subset \R_+$ such that at each location $\xvec{s}_j$ and each $t\in \mathcal{T}$ $Z_i=Z_t(\xvec{s}_i)$ is an element in the simplex $\mathds{S}^{D}$,
\[
\si^{D}=\left\{Z =(Z_{1},Z_{2},\dots ,Z_{D})^{\top}\in \mathds {R}^{D}\,\left|\,Z_{l}\geq 0,l=1,2,\dots ,D;\sum_{l=1}^{D}Z_{l}=\kappa \right.\right\}, 
\]
consisting of the same number of $D$ non-negative mutually dependent components that \textcolor{black}{sum} to a constant $\kappa\in\mathds{R}$. \textcolor{black}{We note that in the original definition of $\si^{D}$, all $D$ parts are strictly assumed to be positive to avoid problems with zeros in log-ratio transformations. More recently, however, \cite{Tsagris2016} introduced so-called $\alpha$-transformations which extend classic log-ratio transformations to work with zeros in some compositional parts}. Consequently, for the specific time instance $t\in \mathcal{T}$ $\xvec{Z}_t$ constitutes an $n \times D$ dimensional matrix that encapsulates the entirety of compositions across all $n$ spatial areal units for the specific time instance $t\in \mathcal{T}$. We note that any vector $\widetilde{Z}_i$ of $D$ real positive components can always be translated into composition-valued information by applying the closure operator $\cls(\widetilde{Z}_i)$ with 
\[
Z_i=\cls(\widetilde{Z}_i)=
\left(\frac{\tilde{Z}_{i1}}{\sum_{l=1}^D \tilde{Z}_{il}},\ldots,\frac{\tilde{Z}_{iD}}{\sum_{l=1}^D \tilde{Z}_{il}}\right)^\top.
\]
Making use of the so-called \textit{Aitchison geometry} \citep{MR1873662}, $\mathds{S}^D$ together with the perturbation and powering operations $\oplus$ and $\odot$ where 
\[
Z_i\oplus Z_j=\cls(Z_{i1}Z_{j1},Z_{i2}Z_{j2},\ldots, Z_{iD}Z_{jD})
\]
and 
\[
\xi\odot Z_i=\cls(Z_{i1}^{\xi},Z_{i2}^{\xi},\ldots, Z_{iD}^{\xi})
\]
 with $Z_i,Z_j\in \si^D$ and $\xi\in\mathds{R}$ can be turned into a Hilbert space with the Aitchison inner product 
\begin{equation}\label{eq:AitInner}
   \langle Z_i, Z_j \rangle_A=\frac{1}{2D}\sum^{D}_{l=1}\sum^D_{k=1} \log\left(\frac{Z_{il}}{Z_{ik}}\right)\log\left(\frac{Z_{jl}}{Z_{jk}}\right), 
\end{equation}
Aitchison norm \textcolor{black}{$\|Z_i\|_A = \sqrt{\langle Z_i, Z_i\rangle_A}$} and the associated Aitchison \textcolor{black}{metric} 
 \[
 d_A(Z_i,Z_j)=\Vert Z_i\ominus Z_j \Vert_A
 \]
where $\Vert Z_i\ominus Z_j \Vert_A=Z_i\oplus((-1)\odot Z_j)$ is the \textcolor{black}{negative perturbation} operation. Instead of working on the simplex $\si^D$, it is often more convenient to work on $\R^{\tilde{D}}$ by applying \textcolor{black}{a} map $\psi:\mathds{S}^D\to \mathds{R}^{\tilde{D}}, Z_i\mapsto \psi(Z_i)$ and performing statistical analysis methods in $\R^{\tilde{D}}$ where ${\tilde{D}}$ is determined by the particular choice of $\psi$ \citep{doi:https://doi.org/10.1002/9781119976462.ch3}. Denoting the $k$-th coordinate of the transformed spatial composition $\psi(Z_i)=(\psi_1(Z_i),\ldots, \psi_{\tilde{D}}(Z_i))^\top$ by $\psi_k(Z_i)$, the underlying idea is to express $Z_i\in \si^D$ in the form of a canonical basis function representation 
 \[
 Z_i=\bigoplus^{\tilde{D}}_{k=1}\psi_k(Z_i)\odot\mathbf{w}_k
 \]
where $\mathbf{w}_k=\cls(\exp(\mathbf{d}_k)),~k=1,\ldots,{\tilde{D}}$, with $\mathbf{d}_k\in\mathds{R}^{\tilde{D}}$ denoting the unit vector associated with the $k$-th coordinate \citep{doi:10.1002/9781119976462}. While different transformations like the \textit{additive log-ratio} (alr) transformation  \citep{10.1093/biomet/67.2.261} have been discussed in the literature, here we only review the centred log-ratio transformation (clr) \citep{10.1093/biomet/70.1.57} and isometric log-ratio transformation\textcolor{black}
{s} (ilr) \citep{Egozcue2003} which establish an isometric isomorphism between $\si^D$ and $\mathds{R}^{\Tilde{D}}$ \citep{BillheimerEtAl2001, PawlowskyGlahn2001}. The recognised isometric isomorphism bridging Aitchison geometry with Euclidean geometry ensures that the inner product in Aitchison space, as well as distances and metrics, corresponds precisely to their Euclidean counterparts once variables have undergone transformation. 

Imposing a sum-to-zero constraint, the clr transformation with \textcolor{black}{$\clr_l(Z_i)=\log(Z_{il}/g(Z_i))$}, where $g(Z_i)$ denotes the geometric mean of the composition $Z_i$, maps the composition from the simplex to a hyperplane $\mathds{H}\subset\mathds{R}^{D}$ that is orthogonal to the vector of ones. Unlike the clr transformation that yields degenerate distributions and singular covariance matrices, the ilr transformation provides a map between $\mathds{S}^D$ and $\mathds{R}^{D-1}$ which corresponds to a class of orthonormal coordinate representations that are derived by applying the Gram-Schmidt procedure to an orthonormal basis $(\mathbf{e}_1, \mathbf{e}_2,\ldots,\mathbf{e}_{D-1})$ on the simplex $\mathds{S}^D$ 
\[
Z_i=\bigoplus_{k=1}^{D-1}\langle Z_i,\mathbf{e}_k\rangle_A\odot\mathbf{e}_k, \ \ 
\]
yielding $\ilr(Z_i) = \left(\langle Z_i,\mathbf{e}_1\rangle_A,\langle Z_i,\mathbf{e}_2\rangle_A,\ldots,\langle Z_i,\mathbf{e}_{D-1}\rangle_A\right)$.
Both the clr and ilr transformations are related to each other through \textcolor{black}{$\ilr(Z_i)=\mathbf{V}_\mathrm{D}\clr(Z_i)$} with $\mathbf{V}_\mathrm{D}$ denotes a $((D-1)\times D)$-dimensional Helmert matrix with rows $\mathbf{v}_j=\clr(\mathbf{e}_j), j=1,\ldots,D-1$  satisfying $\mathbf{V}_\mathrm{D}\mathbf{V}_\mathrm{D}^{\top}=\mathbf{I}_{\mathrm{D}-1}$ and $\mathbf{V}_\mathrm{D}^{\top}\mathbf{V}_\mathrm{D}=\mathbf{I}_{\mathrm{D}}-D^{-1}\mathds{1}_\mathrm{D}\mathds{1}^{\top}_\mathrm{D}$ where $\mathbf{I}_{\mathrm{D}}$ is the identity matrix of dimension $(D\times D)$, and $\mathds{1}_\mathrm{D}$ a $(D\times 1)$ vector of ones. When dealing with a composition comprising of two parts ($D=2$), the $\ilr$ transformation aligns with the logit function typically applied in logistic regression. In scenarios where the number of components exceeds two ($D>2$), a multitude of orthonormal basis systems are possible, all of which substantially influence how projected data is interpreted. Specific configurations of these bases include coordinate representations through \textit{balances} \citep{Egozcue2005:Balances} where each \textit{balancing element} can be understood as the normalised log-ratio of the geometric centres of two defined groups. This concept is rooted in the \textit{sequential binary partition} technique, which entails dividing the composition into two distinct sections. In another configuration called \textit{pivot coordinates} \citep{Fiserova2011, Hron2017}, the very first ilr coefficient mirrors the numerator of the initial clr coefficient, adjusted by a factor of $\sqrt{D/(D-1)}$, rendering it comprehensible as the log-ratio between a specific component and the geometric mean. Conversely, the interpretation of subsequent coefficients proves to be more intricate. To manage this complexity, literature has put forward generalised pivot coordinates, which utilise permuted compositions, alongside symmetric pivot coordinates \citep[see][for further details]{sym:pivot, Hron2021} to aid in better understanding and application.

It is worth noting that the aforementioned formulation necessitates strict positivity across all $D$ compositional elements under investigation. However, this requirement has been addressed by extending both the centred log-ratio (clr) transformation and the isometric log-ratio (ilr) transformation, leading to the development of the centred $\alpha$ ($\aclr$) and isometric $\alpha$ ($\ailr$) transformations. These extensions, as introduced by \citet{Tsagris2016} and further elaborated by \citet{CLAROTTO2022100570}, enable the inclusion of zero values in some of the $D$ components, thereby broadening the applicability of these transformations in practical scenarios.

\subsection{Dynamic spatiotemporal autoregressive panel model}\label{sec:methods}

Consider a compositional spatiotemporal areal process $\xmat{Z}_t$ as described in Section \ref{sec:prelim} carrying information on $D$ compositional part for a discrete set of spatial units comprising $n$ distinct geographical locations $\{\xvec{s}_1, \ldots, \xvec{s}_n\}$ that remains consistent across all temporal instances $t = 1, \ldots, T$. Instead of $\xmat{Z}_t$ and making use of the isometric isomorphism between the simplex and the Euclidean spaces, let $\xmat{Y}_t$ denote the $\psi$-transformed process $\psi(\xmat{Z}_t)$, such that for any temporal instance $t$, $\xvec{Y}_t$ is a $n \times \tilde{D}$-dimensional matrix. In particular, to avoid any degenerative distribution and non-singular covariance matrices, we set $\psi=\ilr$ such that  $\xmat{Y}_t$ is a $n \times (D-1)$-dimensional matrix. Further, we assume that $\xmat{Y}_t$ follows a multivariate spatiotemporal autoregressive process \textcolor{black}{with higher-order temporal lags, i.e.,}
\begin{equation} \label{eq:initial2}
\xmat{Y}_t  =  \sum_{i=1}^{q}\xmat{X}_{t,i}\xvec{\beta}_i + \xmat{W}  \, \xmat{Y}_t \,  \xmat{\Psi}  +  \textcolor{black}{\sum_{\ell = 1}^{L} \xmat{Y}_{t-{\tau_\ell}} \,  \xmat{\Pi}_\ell} + \xmat{E}_t\, ,
\end{equation}
where $\xmat{E}_t = (\xvec{\varepsilon}_{1,t}, \ldots, \xvec{\varepsilon}_{D-1,t})$ is the $n \times (D-1)$-dimensional matrix of disturbances with independent and identically distributed random vectors $\xvec{\varepsilon}_{j,t} = (\varepsilon_{j,t}(\xvec{s}_1), \ldots, \varepsilon_{j,t}(\xvec{s}_n))^\top$ with $\e(\xvec{\varepsilon}_{j,t}) = \xvec{0}$ and $\cov(\xvec{\varepsilon}_{j,t}) = \hat{\sigma}^2 \xmat{I}_n$ for all $j = 1,\ldots, D-1$ and $t = 1,\ldots, T$,~ $\xmat{I}_n$ is the $n$-dimensional identity matrix and  
$\xmat{X}_{t,i}$ is an $n \times (D-1)$-dimensional matrix of the $i$-th exogenous regressors at time point $t$ which enter the regression model with slope coefficients $\xvec{\beta}_i = (\beta_{i,1}, \ldots, \beta_{i,D-1})^\top$. In what follows, we will make use of matrix notation to represent $\xvec{\beta}_i$ in compact form by $\xmat{B} = (\beta_{ij})_{i = 1, ..., q, j = 1,...,D-1}$. Moreover, the model includes a spatial autoregressive term with an $n \times n$-dimensional matrix $\xmat{W}$ of spatial weights \textcolor{black}{and an unknown coefficient matrix $\xmat{\Psi}$ to be estimated}. This matrix defines the spatial proximity structure, i.e., which locations are considered to be adjacent and can thereby influence each other. In practice, this matrix is assumed to be known and determined by the underlying geography. \textcolor{black}{Moreover, the model includes $1 \leq \tau_1 < \ldots < \tau_L$ temporal lags and $\{\xmat{\Pi}_1, \ldots, \xmat{\Pi}_L\}$ are the corresponding $(D-1)$-dimensional square coefficient matrices.} The cross-component spatial effects are represented by the off-diagonal elements of $\xmat{\Psi}$, and the temporally lagged cross-component effects are given by the off-diagonal elements of \textcolor{black}{$\{\xmat{\Pi}_1, \ldots, \xmat{\Pi}_L\}$}. In addition, the component-wise spatial and temporal autoregressive effects are summarised by the diagonal entries of $\xmat{\Psi}$ and \textcolor{black}{$\{\xmat{\Pi}_1, \ldots, \xmat{\Pi}_L\}$}, respectively.

The model will be estimated using a quasi maximum likelihood (QML) approach; combining the results of \cite{otto2024multivariate} on multivariate spatiotemporal GARCH models, \cite{yang2017identification}, who derived identifiability conditions the consistency and asymptotic normality of a QML estimator for multivariate \textcolor{black}{purely} spatial data, and \cite{Yu08}, who derived asymptotic results for a QML estimator for spatiotemporal but univariate processes when both $n$ and $T$ are large. Using the $\vop$-operator for the vectorisation of a matrix, we can rewrite \eqref{eq:initial2} to get the vectorised form
\begin{eqnarray}\label{eq:vec_representation}
\footnotesize{
\vop(\xmat{Y}_t) = 
\vop\left(\sum_{i=1}^{q}\xmat{X}_{t,i}\xvec{\beta}_i\right) 
+ (\xmat{\Psi}^\top \otimes \xmat{W}) \vop(\xmat{Y}_t) 
+ \textcolor{black}{\sum_{\ell=1}^{L}(\xmat{\Pi}_\ell^\top \otimes \xmat{I}_n) \vop(\xmat{Y}_{t-\tau_\ell})
+ \vop(\xmat{E}_t)\, }.
}
\end{eqnarray}
The Kronecker product is denoted by $\otimes$. Interestingly, using such vec-representation, one can see that the multivariate spatiotemporal autoregressive model with $n$ spatial units is a special case of a (univariate) $n(D-1)$-dimensional spatiotemporal autoregressive model with a weight matrix $\xmat{\Psi}^\top \otimes \xmat{W}$ (i.e., having $n(D-1)$ artificial spatial units).

\textcolor{black}{For an easier notation, let $\ddot{\xvec{Y}}_t = \vop(\xmat{Y}_t)$, $\xvec{u}_t = \vop(\xmat{E}_t)$,
$\xmat{A}_\ell(\xmat{\Pi}_\ell)=\xmat{\Pi}_\ell^\top\otimes \xmat{I}_n$,
$\xmat{S}_{n(D-1)}(\xmat{\Psi}) = \xmat{I}_{n(D-1)} - \xmat{\Psi}^\top\otimes \xmat{W}_n$, and
$\xvec{x}_t(\xmat{B}) = \vop\!\Big(\sum_{i=1}^q \xmat{X}_{t,i}\xvec{\beta}_i\Big)$.
Then, we can express the vectorised model in reduced form as
\begin{equation}\label{eq:reduced_cons}
\ddot{\xvec{Y}}_t
=
\xmat{S}^{-1}_{n(D-1)}(\xmat{\Psi})\,  \left( \xvec{x}_t(\xmat{B})
+ \sum_{\ell=1}^{L}\xmat{A}_\ell(\xmat{\Pi}_\ell)\,\ddot{\xvec{Y}}_{t-\tau_\ell}
+ \xvec{u}_t \right),
\end{equation}
which provides the residual map for
$\vartheta=(\vop(\xmat{B})^\top, \vop(\xmat{\Psi})^\top, \vop(\xmat{\Pi}_1)^\top,\ldots,\vop(\xmat{\Pi}_L)^\top)^\top$,
\[
\xvec{r}_t(\vartheta)
=
\xmat{S}_{n(D-1)}(\xmat{\Psi}) \ddot{\xvec{Y}}_{t}
-\xvec{x}_t(\xmat{B})
-\sum_{\ell=1}^{L}\xmat{A}_\ell(\xmat{\Pi}_\ell)\,\ddot{\xvec{Y}}_{t-\tau_\ell}.
\]
Recall $\xmat{S}_{n(D-1)}= \xmat{I}_{n(D-1)}- \xmat{\Psi} ^\top\!\otimes \xmat{W}$.  A convenient sufficient condition for invertibility is $\rho(\xmat{\Psi}^\top\!\otimes \xmat{W})<1$, where $\rho(\cdot)$ denotes the spectral radius. Using $\rho(\xmat{A}\otimes \xmat{B})=\rho(\xmat{A})\rho(\xmat{B})$, this is satisfied whenever  $\rho(\xmat{\Psi})\rho(\xmat{W})<1$. For row-standardised spatial weight matrices, $\rho(\xmat{W})\le 1$, implying that $\rho(\xmat{\Psi})<1$ ensures existence of the spatial multiplier $\xmat{S}_{n(D-1)}^{-1}=\sum_{k=0}^{\infty}(\xmat{\Psi}^\top\!\otimes \xmat{W})^k$. In the empirical applications, we verify this condition numerically by checking that the eigenvalues of $\xmat{S}_{n(D-1)}$ are bounded away from zero.}

\textcolor{black}{Under the data-generating parameter \linebreak $\vartheta_0 = (\vop(\xmat{B}_0)^\top, \vop(\xmat{\Psi}_0)^\top, \vop(\xmat{\Pi}_{1,0})^\top, \ldots,  \vop(\xmat{\Pi}_{\ell,0})^\top)^\top$, we have $\xvec{r}_t(\vartheta_0)=\xvec{u}_t$.  Let $\tau_{\max} = \max_{\ell=1,\ldots,L}\tau_\ell$ and $T^\star = T-\tau_{\max}$ and $\sigma^2 > 0$ denote the (unknown) innovation variance parameter.
The Gaussian quasi log-likelihood (up to additive constants independent of $(\vartheta,\sigma^2)$) is given by
\small{
\begin{equation}\label{eq:qll_cons}
\log \mathcal{L}_{nT}(\vartheta,\sigma^2)
=
T^\star\log|\xmat{S}_{n(D-1)}(\xmat{\Psi})|
-\frac{T^\star n(D-1)}{2}\log\sigma^2
-\frac{1}{2\sigma^2}\sum_{t=\tau_{\max}+1}^{T} \xvec{r}_t(\vartheta)^\top \xvec{r}_t(\vartheta).
\end{equation}
}
Let $(\hat\vartheta_{nT},\hat\sigma^2_{nT})$ maximise \eqref{eq:qll_cons} over the parameter space
$\Theta\times[\underline\sigma^2,\overline\sigma^2]$, where
$0<\underline\sigma^2<\sigma_0^2<\overline\sigma^2<\infty$.}

To establish the consistency of the quasi-maximum likelihood (QML) estimator for our spatiotemporal autoregressive compositional panel data model, we require a set of assumptions. These assumptions govern the behaviour of the error process, the parameter space, some standard assumptions on the spatial weight matrix, and the asymptotic framework under which consistency can be established.

\begin{assumption}\label{ass:err_exo_cons}
\textcolor{black}{Assume that $\{\xvec{u}_t\}_{t\ge1}$ is i.i.d.\ with $\E(\xvec{u}_t)=\xvec{0}$ and $\Var(\xvec{u}_t)=\sigma_0^2 \xmat{I}_{n(D-1)}$. Furthermore, let $\mathcal{F}_{t-1}=\sigma\{\xvec{u}_s,\xmat{X}_s: s\le t-1\}$ and
$\mathcal{G}_t=\sigma(\mathcal{F}_{t-1},\xmat{X}_t)$. Assume strict exogeneity, i.e.\
$\E(\xvec{u}_t\mid \mathcal{G}_t)=\xvec{0}$.
Finally, there exists $\eta > 0$ such that $\sup_{n \geq 1}\E\!\left(\|\xvec{u}_t\|^{4+\eta}\right) < \infty.$}
\end{assumption}

Assumption \ref{ass:err_exo_cons} ensures that \textcolor{black}{innovation process has well-defined, finite second and higher-order moments and is serially independent over time. The finite $(4+\eta)$-moment condition ensures that quadratic forms of the residuals satisfy a law of large numbers and a central limit theorem, which is required for consistency and asymptotic normality of the QML estimator.} From a practical perspective, \textcolor{black}{this assumption is satisfied} in many empirical applications where extreme shocks are \textcolor{black}{rare}. In case of many outliers or (extremely) heavy-tailed distributions, the estimation results \textcolor{black}{and inference} could be distorted. \textcolor{black}{If the regressors are stochastic rather than deterministic, the following additional assumption is required; in the case of deterministic and uniformly bounded regressors, this assumption can be omitted.}

\begin{assumption}\label{ass:X_stoch_cons}
\textcolor{black}{Assume that:
\begin{enumerate}
\item[(i)] $\{\xmat{X}_t\}_{t\ge1}$ is strictly stationary and ergodic, where
$\xmat{X}_t=(\xmat{X}_{t,1},\ldots,\xmat{X}_{t,q})$;
\item[(ii)] $\E(\|\xmat{X}_t\|^{4+\eta})<\infty$ for the same $\eta$ as in
Assumption~\ref{ass:err_exo_cons}.
\end{enumerate}
}
\end{assumption}

\textcolor{black}{To ensure that the spatiotemporal autoregressive system in \eqref{eq:reduced_cons} is well-defined and admits a stable causal solution (uniformly over the admissible parameters),} we impose the following stability and compactness condition; \textcolor{black}{the innovation variance is restricted to $\sigma^2\in[\underline\sigma^2,\overline\sigma^2]$ as in \eqref{eq:qll_cons}.} 

\textcolor{black}{For $r=1,\ldots,\tau_{\max}$, let
\[
\xmat{\Phi}_r(\vartheta)=
\begin{cases}
\xmat{S}_{n(D-1)}(\xmat{\Psi})^{-1}
(\xmat{\Pi}_\ell^\top\otimes\xmat{I}_n),
& \text{if } r=\tau_\ell \text{ for some } \ell,\\[1mm]
\xmat{0}_{n(D-1)\times n(D-1)}, & \text{otherwise}.
\end{cases}
\]
Further, the homogeneous part of the reduced-form dynamics can be written as
\[
\ddot{\xvec{Y}}_t=\sum_{r=1}^{\tau_{\max}}\xmat{\Phi}_r(\vartheta)\ddot{\xvec{Y}}_{t-r}.
\]
Let $\xvec{Z}_t= (\ddot{\xvec{Y}}_t^\top, \ddot{\xvec{Y}}_{t-1}^\top, \ldots, \ddot{\xvec{Y}}_{t-\tau_{\max}+1}^\top)^\top$.
Then, the recursion can be written as
\[
\xvec{Z}_t=\xmat{F}(\vartheta)\xvec{Z}_{t-1},
\]
where $\xmat{F}(\vartheta)$ is the block companion matrix
\[
\xmat{F}(\vartheta)=
\begin{pmatrix}
\xmat{\Phi}_1(\vartheta) & \cdots & \xmat{\Phi}_{\tau_{\max}-1}(\vartheta)& \xmat{\Phi}_{\tau_{\max}}(\vartheta)\\
\xmat{I}_m & \cdots & \xmat{0} & \xmat{0}\\
\vdots & \ddots & \vdots & \vdots\\
\xmat{0} & \cdots & \xmat{I}_m & \xmat{0}
\end{pmatrix}.
\]
}

\begin{assumption}\label{ass:stability_cons}
\textcolor{black}{
For all $\vartheta=(\vop(\xmat{B})^\top,\vop(\xmat{\Psi})^\top,\vop(\xmat{\Pi})^\top)^\top$ in a compact parameter space $\Theta$,
the matrices $\xmat{S}_{n(D-1)}(\xmat{\Psi})$ are invertible and satisfy
\[
\sup_{\vartheta\in\Theta,n}\|\xmat{S}_{n(D-1)}(\xmat{\Psi})^{-1}\|_1<\infty,
\qquad
\sup_{\vartheta\in\Theta,n}\|\xmat{S}_{n(D-1)}(\xmat{\Psi})^{-1}\|_\infty<\infty .
\]
Further, the reduced-form time dynamics are uniformly stable. In particular,
there exists $\rho\in(0,1)$ such that
\[
\sup_{\vartheta\in\Theta,n}
\rho\!\big(\xmat{F}(\vartheta)\big)<\rho .
\]}
\end{assumption}

Since our model includes a spatial autoregressive component, we must also impose regularity conditions on the spatial weight matrix \( \xmat{W} \), as formulated in Assumption \ref{ass:W_cons}.  

\begin{assumption}\label{ass:W_cons}
The spatial weights matrix $\xmat{W} = \xmat{W}_n$ is non-stochastic with zero diagonal and uniformly bounded absolute row and column sums in $n$. \textcolor{black}{Moreover, $\xmat{W} \neq 0$ (at least one non-zero element).}
\end{assumption}

These conditions ensure that spatial dependence remains well-behaved as $n$ increases, thereby maintaining the stability of the estimation procedure. In practical applications, this assumption is typically met with standard choices of the spatial weight matrix, such as those based on geographic adjacency or distance-based kernel functions, which naturally enforce decay in spatial interactions. This prevents scenarios where each location has an unbounded number of influential neighbours, which could otherwise result in singularities in the estimation process. \textcolor{black}{To rule out collinearity between regressors and lagged responses and to ensure identification of $(\xmat{B},\xmat{\Pi})$, we impose the following full-rank condition.}

\begin{assumption}\label{ass:rank_id_cons}
\textcolor{black}{Assume $\E(z_t z_t^\top)$ exists and is positive definite, where
\[
z_t=\Big(\vop(\xmat{X}_t)^\top,\ \ddot{\xvec{Y}}_{t-\tau_1}^\top,\ldots,\ddot{\xvec{Y}}_{t-\tau_L}^\top\Big)^\top .
\]}
\end{assumption}

Finally, we need an appropriate asymptotic framework to derive the consistency result, as provided in Assumption \ref{ass:asymptotics_cons}. 

 \begin{assumption}\label{ass:asymptotics_cons}
	Let $n$ be a non-decreasing function of $T$ and $T \to \infty$.
\end{assumption}

Given the above assumptions, we can now establish the main theoretical result showing the consistency of the QML estimator in Theorem \ref{th:consistency}. This result ensures that our estimates converge to the true parameter values as the length of the panel increases, providing the foundation for inference in the spatiotemporal autoregressive compositional panel model. The imposed assumptions collectively guarantee that the estimation problem is well-posed, the likelihood function behaves regularly, and the model remains stable under increasing sample sizes, leading to a consistent QML estimator.

\begin{theorem}\label{th:consistency}
Under Assumptions~\ref{ass:err_exo_cons} to \ref{ass:asymptotics_cons}, the parameters \textcolor{black}{$(\vartheta_0,\sigma_0^2)$} are uniquely identifiable and the \textcolor{black}{Gaussian} QML estimator is a consistent estimator
\[
\hat\vartheta_{nT}\xrightarrow{p}\vartheta_0,\qquad
\textcolor{black}{\hat\sigma^2_{nT}\xrightarrow{p}\sigma_0^2,}
\]
for $T \to \infty$.
\end{theorem}

The proof of the Theorem is provided in the Appendix. \textcolor{black}{While Theorem~\ref{th:consistency} establishes convergence in probability, asymptotic normality of $(\hat\vartheta_{nT},\hat\sigma^2_{nT})$ requires additional smoothness and moment conditions ensuring that a Taylor expansion of the score is valid and that the (normalised) score satisfies a central limit theorem under the joint asymptotic regime $T\to\infty$ with $n=n(T)$ non-decreasing. For that reason, we need some additional growth constraints on $n$.}

\textcolor{black}{
\begin{assumption}\label{ass:AN_reg}
Assume:
\begin{enumerate}
\item[(i)] (\emph{Non-singular information})
The matrix
\[
\mathcal{J}
=
-\E\!\left[\nabla^2_{(\vartheta,\sigma^2)}\ell_t(\vartheta_0,\sigma_0^2)\right]
\]
exists, is finite, and positive definite.
\item[(ii)] (\emph{Central limit theorem for the score})
The score satisfies
\[
\frac{1}{\sqrt{T}}\sum_{t=1}^T
\nabla_{(\vartheta,\sigma^2)}\ell_t(\vartheta_0,\sigma_0^2)
\xrightarrow{d} N(0,\mathcal{I}),
\]
with $\mathcal{I}$ finite.
Under Assumption~\ref{ass:err_exo_cons},
$\mathcal{I}=\Var\!\big(\nabla_{(\vartheta,\sigma^2)}\ell_t(\vartheta_0,\sigma_0^2)\big)$.
\item[(iii)] (\emph{Joint asymptotics})
The temporal dimension satisfies $T\to\infty$, while the spatial
dimension may remain fixed or increase with $T$, i.e.\ $n=n(T)$.
If $n(T)\to\infty$, assume the mild growth restriction
\[
\frac{n(T)}{T}\to0,
\]
and that the moment and stability bounds in
Assumptions~\ref{ass:err_exo_cons} and~\ref{ass:stability_cons}
hold uniformly in $n$.
\end{enumerate}
\end{assumption}
}

\textcolor{black}{
Assumption~\ref{ass:AN_reg} introduces only the additional conditions required to strengthen consistency to asymptotic normality. Smoothness of the Gaussian quasi log-likelihood and local uniform boundedness of its derivatives follow from the structural properties of the model and the earlier assumptions. In particular, the residual map is affine in the parameters and the log-determinant term is smooth whenever $\xmat{S}_{n(D-1)}(\xmat{\Psi})$ is invertible, which holds under Assumption~\ref{ass:stability_cons}. Combined with compactness of the parameter space and the moment bounds on the innovations, this implies that the likelihood is twice continuously differentiable in a neighbourhood of $(\vartheta_0,\sigma_0^2)$ and that its Hessian converges locally uniformly in probability to its expectation.
The asymptotic theory is therefore driven by temporal aggregation. The temporal dimension satisfies $T\to\infty$, while the spatial dimension may remain fixed or increase with $T$. The mild growth restriction $n(T)/T\to0$ ensures that the cross-sectional dimension does not expand too quickly relative to the temporal domain. Since each single-period likelihood contribution is already normalised by the cross-sectional dimension $n(D-1)$, this condition guarantees that the score satisfies a central limit theorem and that the Hessian can be replaced by its population counterpart, yielding the usual $\sqrt{T}$-rate for the quasi-maximum likelihood estimator.
}

\textcolor{black}{
\begin{theorem}\label{thm:AN}
Under Assumptions~\ref{ass:err_exo_cons}--\ref{ass:asymptotics_cons}
and \ref{ass:AN_reg},
\[
\sqrt{T}\,\big((\hat\vartheta_{nT},\hat\sigma^2_{nT})-(\vartheta_0,\sigma_0^2)\big)
\xrightarrow{d}
N\!\left(\xvec{0},\,\mathcal{J}^{-1}\mathcal{I}\,\mathcal{J}^{-1}\right),
\]
where
\[
\mathcal{J}
=
-\E\!\left[\nabla^2_{(\vartheta,\sigma^2)}\ell_t(\vartheta_0,\sigma_0^2)\right],
\qquad
\mathcal{I}
=
\sum_{h=-\infty}^{\infty}\Cov\!\big(
\nabla_{(\vartheta,\sigma^2)}\ell_t(\vartheta_0,\sigma_0^2),
\nabla_{(\vartheta,\sigma^2)}\ell_{t-h}(\vartheta_0,\sigma_0^2)
\big).
\]
If, in addition, the Gaussian likelihood is correctly specified, i.e.
$\{\xvec{u}_t\}_{t\ge1}$ are i.i.d.\ $N(\xvec{0},\sigma_0^2\xmat{I}_{n(D-1)})$,
then the information identity holds and $\mathcal{I}=\mathcal{J}$, yielding
\[
\sqrt{T}\,\big((\hat\vartheta_{nT},\hat\sigma^2_{nT})-(\vartheta_0,\sigma_0^2)\big)
\xrightarrow{d}
N\!\left(\xvec{0},\,\mathcal{J}^{-1}\right).
\]
\end{theorem}
}

\textcolor{black}{The proof of the theorem is provided in the Appendix. As a consequence of Theorem~\ref{thm:AN}, asymptotically valid} standard errors can be obtained from the \textcolor{black}{estimated asymptotic covariance matrix. In general, this is given by the sandwich form $\mathcal{J}^{-1}\mathcal{I}\mathcal{J}^{-1}$, where $\mathcal{J}$ and $\mathcal{I}$ are consistently estimated by their sample analogues. If the Gaussian likelihood is correctly specified, the information identity holds and the covariance matrix simplifies to $\mathcal{J}^{-1}$, which can be estimated} by the inverse of the observed Hessian of the log-likelihood evaluated at the QML estimator.

\textcolor{black}{In the present spatiotemporal panel setting, the quasi log-likelihood can be written as
\[
\ell_{nT}(\vartheta,\sigma^2)
=
\frac{1}{T}\sum_{t=1}^T \ell_t(\vartheta,\sigma^2),
\]
where $\ell_t$ denotes the normalised contribution of time point $t$.  Under correct specification of the temporal dynamics and Assumption~\ref{ass:err_exo_cons},  the score contributions  $\nabla_{(\vartheta,\sigma^2)} \ell_t(\vartheta_0,\sigma_0^2)$  form a martingale difference sequence and are serially uncorrelated across $t$.  In this case, the matrix $\mathcal{J}$ can be consistently estimated by the negative time-average of the observed Hessian,
\[
\widehat{\mathcal{J}}
=
-
\frac{1}{T}
\sum_{t=1}^T
\nabla^2_{(\vartheta,\sigma^2)} 
\ell_t(\hat\vartheta_{nT},\hat\sigma^2_{nT}),
\]
and $\mathcal{I}$ by the sample covariance of the score contributions,
\[
\widehat{\mathcal{I}}
=
\frac{1}{T}
\sum_{t=1}^T
\hat{s}_t \hat{s}_t^\top,
\qquad
\hat{s}_t
=
\nabla_{(\vartheta,\sigma^2)}
\ell_t(\hat\vartheta_{nT},\hat\sigma^2_{nT}).
\]
}

\textcolor{black}{If the temporal dependence is not fully captured by the specified lag structure,  the score process may exhibit residual serial correlation.  In that case, $\mathcal{I}$ should be estimated using a long-run variance estimator,  for example a heteroscedasticity and autocorrelation consistent (HAC) estimator with suitable kernel and bandwidth.  Such estimation requires a sufficiently long time dimension $T$. When the dynamic specification is adequate and $T$ is moderate, the simpler time-averaged Hessian and score covariance estimators are typically preferred.}

\subsection{\textcolor{black}{Model interpretation and effects on the simplex}}

\textcolor{black}{The parameter matrices $\xmat{\Psi}$ and $\xmat{\Pi}$ describe linear spatiotemporal dependence in the transformed Euclidean space induced by the mapping $\psi:\mathds{S}^D \rightarrow \mathds{R}^{\tilde D}$. For $\psi=\ilr$, we have $\tilde D = D-1$. Due to the non-linear nature of the inverse transformation $Z=\psi^{-1}(Y)$, these parameters do not correspond to constant linear effects on the original compositional scale. Instead, the implied dynamics on the simplex are inherently non-linear and state dependent.}

\textcolor{black}{Interpretation on the compositional scale is therefore based on local marginal effects. Let $Z_t(\xvec{s}_i)=\psi^{-1}(Y_t(\xvec{s}_i))$ denote a reference composition at spatial unit $\xvec{s}_i$ and time $t$, and define the Jacobian matrix of the inverse transformation as
\[
\mathbf{J}_\psi\!\left(Y_t(\xvec{s}_i)\right)
=
\frac{\partial \psi^{-1}(Y_t(\xvec{s}_i))}{\partial Y_t(\xvec{s}_i)^\top}
\in \mathds{R}^{D\times (D-1)}.
\]
Conditional on a reference composition $Z_t(\xvec{s}_i)$, local effects of past outcomes are characterised lag-wise. Specifically, the local marginal effect of $Y_{t-\tau_\ell}(\xvec{s}_i)$ on $Z_t(\xvec{s}_i)=\psi^{-1}(Y_t(\xvec{s}_i))$ is governed by $\mathbf{J}_\psi(Y_t(\xvec{s}_i))\,\xmat{\Pi}_\ell^\top$, where the Jacobian is evaluated at the contemporaneous state $Y_t(\xvec{s}_i)$ and $\xmat{\Pi}_\ell$ governs the propagation of past innovations from $t-\tau_\ell$ to $t$.}

\textcolor{black}{Contemporaneous spatial dependence propagates through the reduced form representation
\[
\small{\mathrm{vec}(\xmat{Y}_t)
=
\left(\xmat{I}_{n(D-1)}-\xmat{\Psi}^\top\!\otimes \xmat{W}\right)^{-1}
\left[
\vop\!\left(\sum_{i=1}^{q}\xmat{X}_{t,i}\xvec{\beta}_i\right)
+
\sum_{\ell=1}^{L}(\xmat{\Pi}_\ell^\top\otimes \xmat{I}_n)\,\mathrm{vec}(\xmat{Y}_{t-\tau_\ell})
+
\vop(\xmat{E}_t)
\right],}
\]
where the inverse matrix captures the spatial transmission of shocks across units and components. Accordingly, the impact of a perturbation at spatial unit $\xvec{s}_j$ on unit $\xvec{s}_i$ is governed by the $(i,j)$ block of the reduced form matrix, and its effect on the original compositions is obtained by post-multiplication with $\mathbf{J}_\psi(Y_t(\xvec{s}_i))$.}

\textcolor{black}{This construction yields natural definitions of direct effects (own-unit impacts, $i=j$) and indirect or spillover effects (cross-unit impacts, $i\neq j$) on the simplex. All effects are evaluated locally at meaningful reference compositions and are expressed in terms of changes in compositional shares. While the parameter matrices $\xmat{\Psi}$ and $\xmat{\Pi}$ depend on the chosen ilr basis, the induced effects on the simplex are invariant under orthonormal reparametrisations, providing an interpretable and basis-independent link between the linear spatiotemporal model in transformed space and the non-linear geometry of composition-valued data. Similar arguments for emphasising simplex-intrinsic interpretations rather than coordinate-dependent coefficients have recently been discussed in the compositional regression literature \citep{DARGEL2024107945}.}

\textcolor{black}{The interpretation of the regression coefficients $\xmat{B}$ associated with the exogenous covariates follows similar principles. Since the covariates enter the model linearly in ilr-transformed space, each coefficient $\beta_{i,k}$ represents a constant marginal effect of the corresponding regressor on the $k$-th ilr coordinate of the composition in the structural form of the model.} \textcolor{black}{However, due to the presence of contemporaneous spatial dependence, covariate effects propagate through space according to the reduced form of the model. In particular, a change in an exogenous regressor at spatial unit $\xvec{s}_j$ affects $\xmat{Y}_t$ through the spatial multiplier
\[
\left(\xmat{I}_{n(D-1)}-\xmat{\Psi}^\top\!\otimes \xmat{W}\right)^{-1},
\]
giving rise to both direct effects (on the same unit) and indirect or spillover effects (on other units).} \textcolor{black}{On the original compositional scale, the marginal effect of a covariate is obtained by mapping the reduced-form impact back to the simplex via the inverse ilr transformation. Locally, at a reference composition $Z_t(\xvec{s}_i)$, the effect of a covariate perturbation is therefore characterised by
\[
\mathbf{J}_\psi(Y_t(\xvec{s}_i))
\left[
\left(\xmat{I}_{n(D-1)}-\xmat{\Psi}^\top\!\otimes \xmat{W}\right)^{-1}
\vop(\xmat{X}_{t,i}\xvec{\beta}_i)
\right],
\]
yielding changes in compositional shares that respect the unit-sum constraint. This construction allows covariate effects to be decomposed into direct and indirect components on the simplex, in close analogy to effect decomposition in spatial econometric models, while accounting for the non-linear geometry of compositional data induced by the inverse log-ratio transformation.}

\section{Monte-Carlo Simulations}\label{sec:sim}

In the following section, we will present the results obtained from a series of simulations assessing the \textcolor{black}{performance of the estimators} in finite sample sizes. Therefore, we simulated a spatiotemporal model for 9 different sample sizes with an increasing size of the spatial field $n \in \{4^2, 6^2, 8^2\}$ and the length of the time series $T \in \{30, 80, 160\}$. All processes were simulated on a quadratic two-dimensional grid $\{\xset{Z}^2: 0 < s_1, s_2 \leq \sqrt{n}\}$ and all cells were considered to be adjacent if they share a common border, i.e., $\xmat{W}$ was considered to be a Queen's contiguity matrix. Moreover, the weight matrix was row-standardised, and the simulation study was conducted with \textcolor{black}{$1064$} replications. 
\textcolor{black}{In each replication, covariates (including an intercept) are generated independently and innovations are drawn either from a Gaussian distribution or from a Gaussian mixture (0.95 of a standard normal and 0.05 of a zero-mean Gaussian with variance 25) to induce deviations from normality. The regression coefficient matrix is fixed across settings as $\mathbf B=\bigl((1,2)^\top,(-2,1)^\top,(3,-2)^\top\bigr)^\top$ (i.e., $k=3$ and $p=2$). Moreover, we considered two different settings for the degree of spatiotemporal dependence.}

\textcolor{black}{
\emph{Setting A (moderate overall dependence).} The data-generating process uses a single temporal lag, $\tau_{\text{dgp}}=\{1\}$. Instantaneous spatial dependence is strong and diagonal,
\[
\mathbf\Psi=
\begin{pmatrix}
0.7 & 0\\
0 & 0.7
\end{pmatrix},
\qquad
\mathbf\Pi_{1}=
\begin{pmatrix}
0.2 & 0.1\\
0.0 & 0.1
\end{pmatrix}.
\]
For the lattice sizes considered, the resulting spatiotemporal transition operator has spectral radius of about $0.6$, implying moderate persistence and a comparatively well-conditioned estimation problem. For setting A, we included two shorter time horizons of 5 and 10 to assess the performance for very short time series.}

\textcolor{black}{
\emph{Setting B (high persistence, multi-lag temporal dynamics).} The DGP includes two temporal lags, $\tau_{\text{dgp}}=\{1,12\}$. Spatial dependence is slightly weaker but includes cross-coordinate effects,
\[
\mathbf\Psi=
\begin{pmatrix}
0.5 & 0.1\\
0.2 & 0.5
\end{pmatrix},
\qquad
\mathbf\Pi_{1}=
\begin{pmatrix}
0.1 & 0.2\\
0.1 & 0.1
\end{pmatrix},
\qquad
\mathbf\Pi_{12}=0.3\,\mathbf\Pi_{1}.
\]
Here the implied transition operator has spectral radius close to $0.9$, producing substantially higher persistence and slower effective information accumulation in $T$. This setting is designed to stress finite-sample inference, in particular standard error estimation and coverage under both Gaussian and non-Gaussian innovations. The smallest time series length for setting B was chosen to be 30, leading to an effective length of 18 time points when removing the first 12 time points due to the multi-lag structure.}

\begin{figure}[t]
\centering
\begin{subfigure}[t]{0.48\textwidth}
    \centering
    \includegraphics[width=\linewidth]{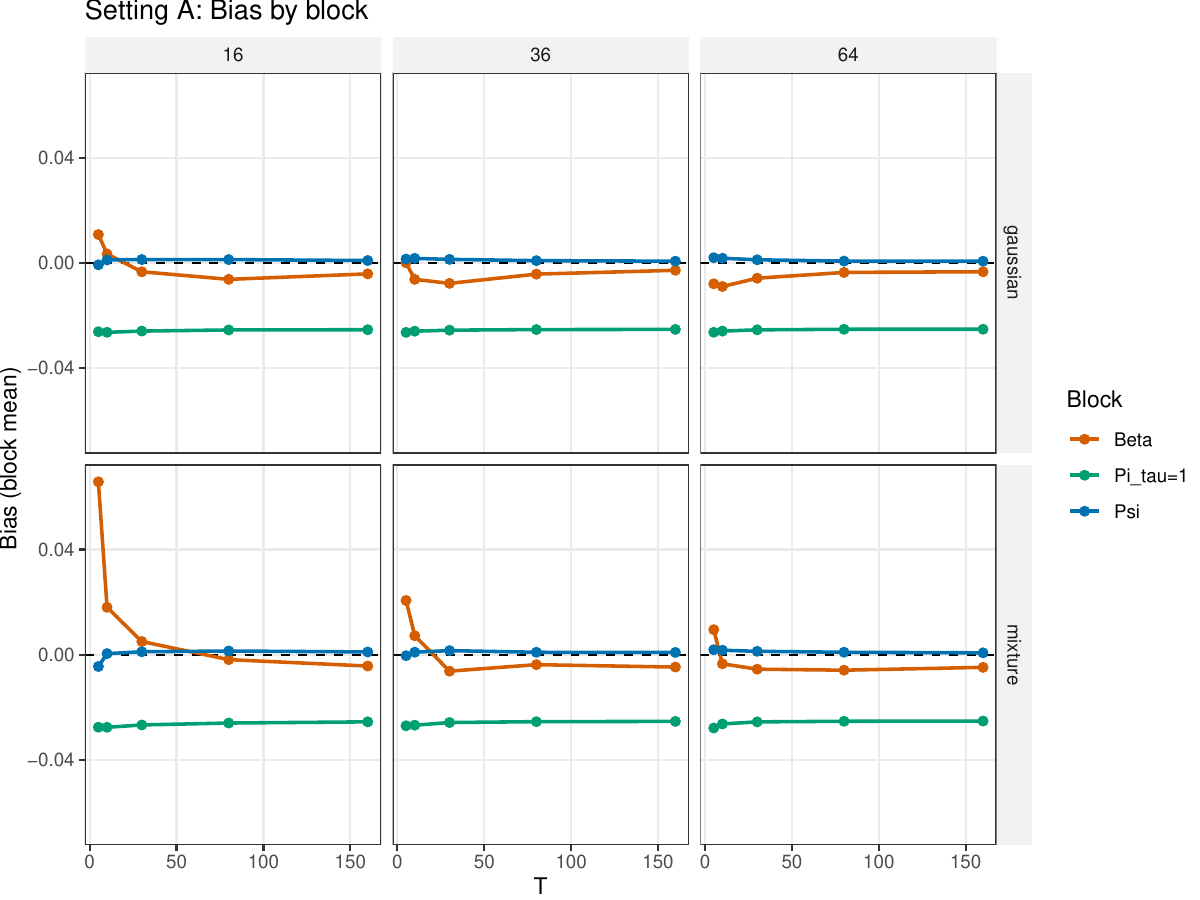}
    \caption{Setting A}
\end{subfigure}
\hfill
\begin{subfigure}[t]{0.48\textwidth}
    \centering
    \includegraphics[width=\linewidth]{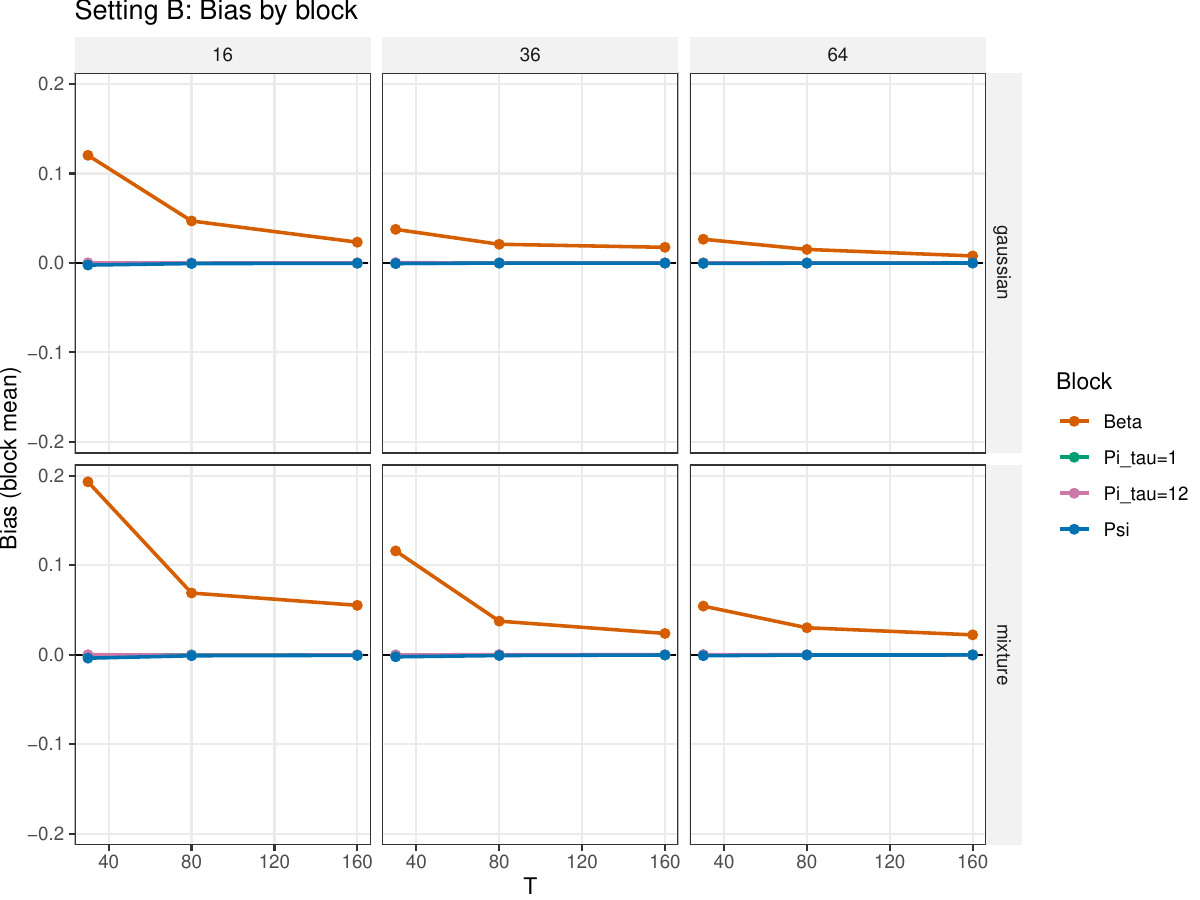}
    \caption{Setting B}
\end{subfigure}
\caption{Finite-sample bias under strong spatial dependence (setting A, left) and multi-lag dependence (setting B, right).}
\label{fig:bias_sim}
\end{figure}

\begin{figure}[t]
\centering
\begin{subfigure}[t]{0.48\textwidth}
    \centering
    \includegraphics[width=\linewidth]{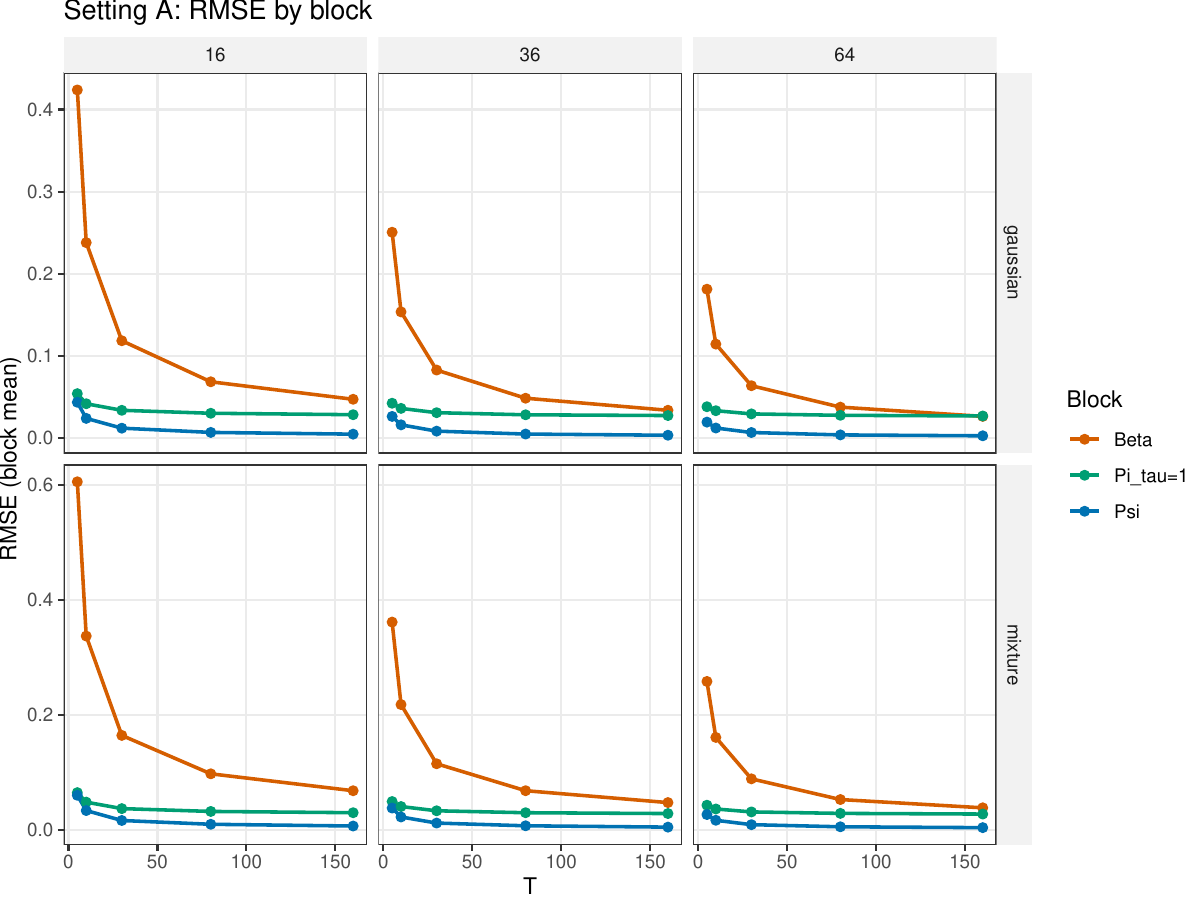}
    \caption{Setting A}
\end{subfigure}
\hfill
\begin{subfigure}[t]{0.48\textwidth}
    \centering
    \includegraphics[width=\linewidth]{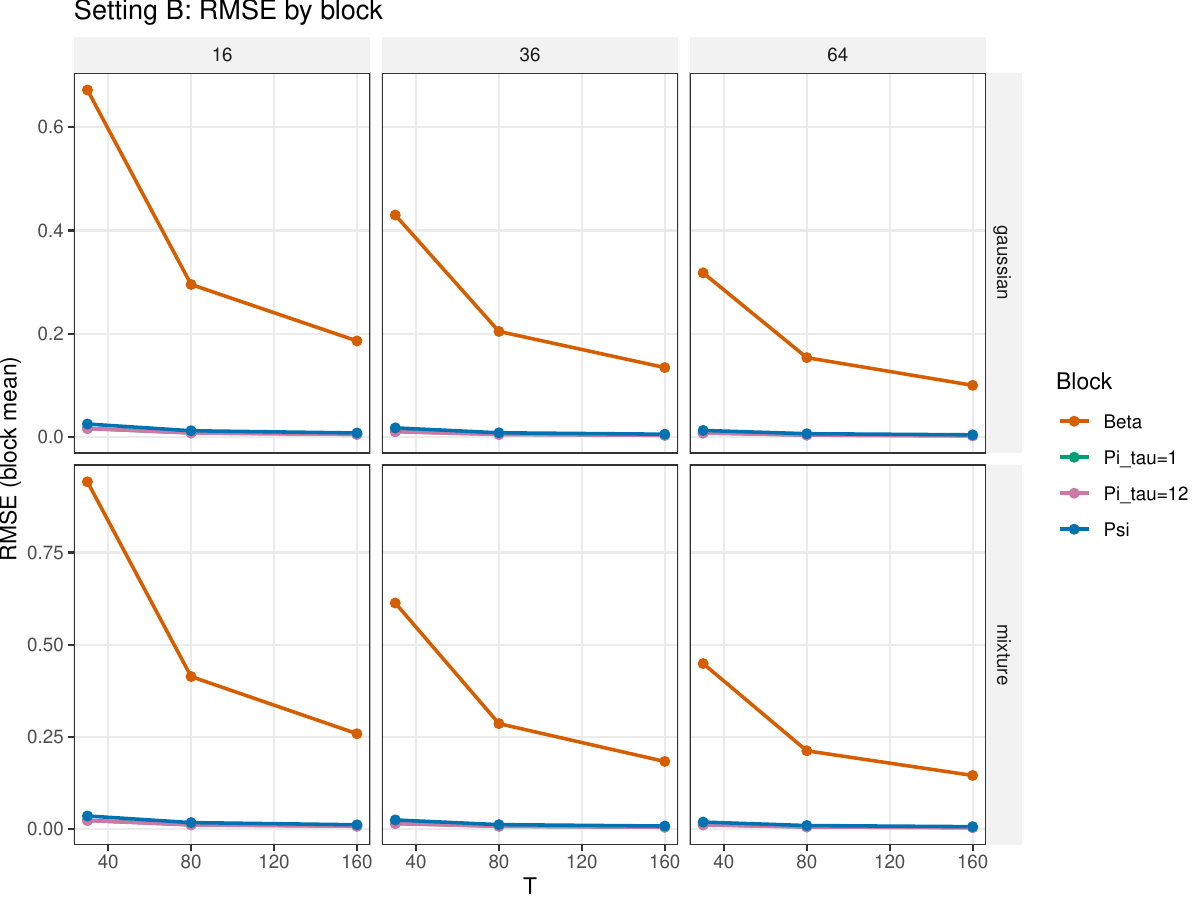}
    \caption{Setting B}
\end{subfigure}
\caption{Relative estimation efficiency across sample sizes for settings A (left) and B (right).}
\label{fig:rmse_sim}
\end{figure}

\begin{figure}[t]
\centering
\begin{subfigure}[t]{0.48\textwidth}
    \centering
    \includegraphics[width=\linewidth]{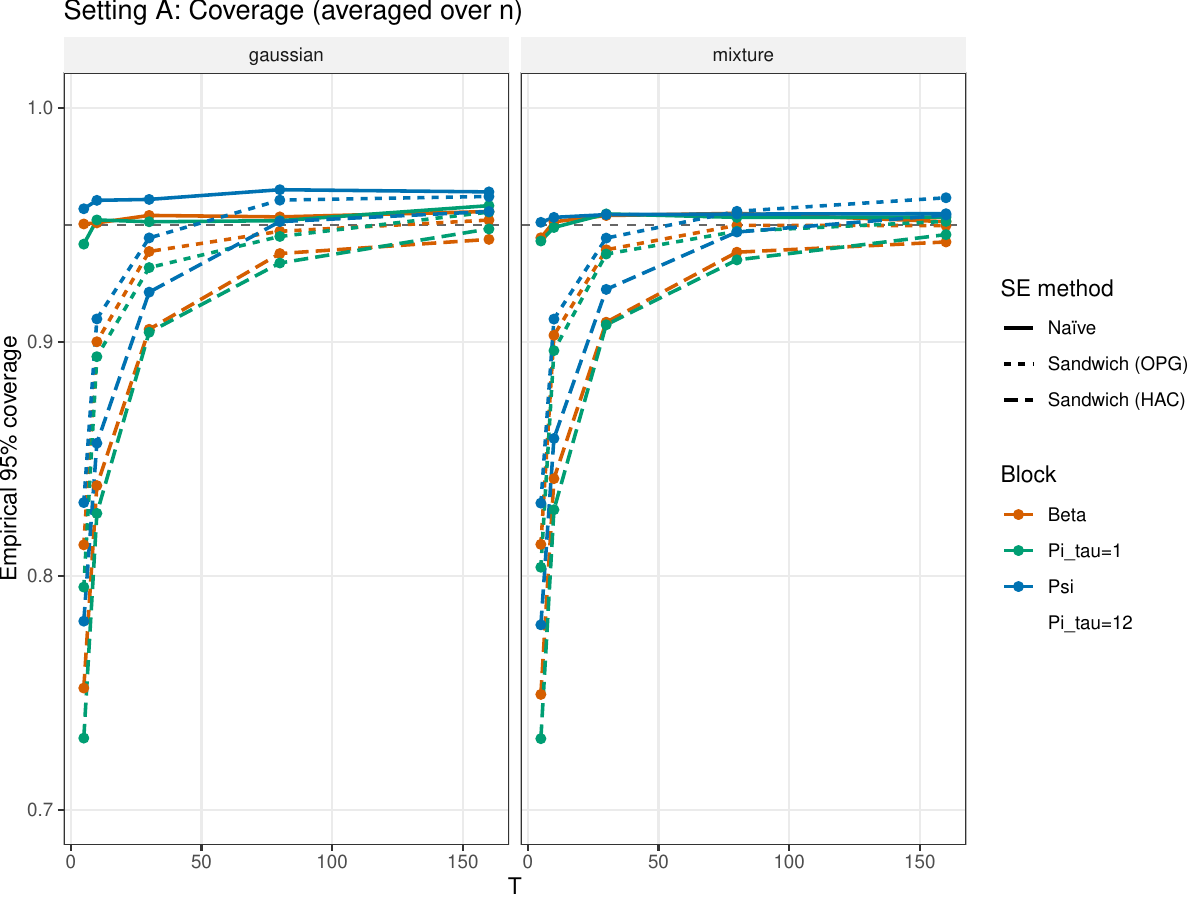}
    \caption{Setting A}
\end{subfigure}
\hfill
\begin{subfigure}[t]{0.48\textwidth}
    \centering
    \includegraphics[width=\linewidth]{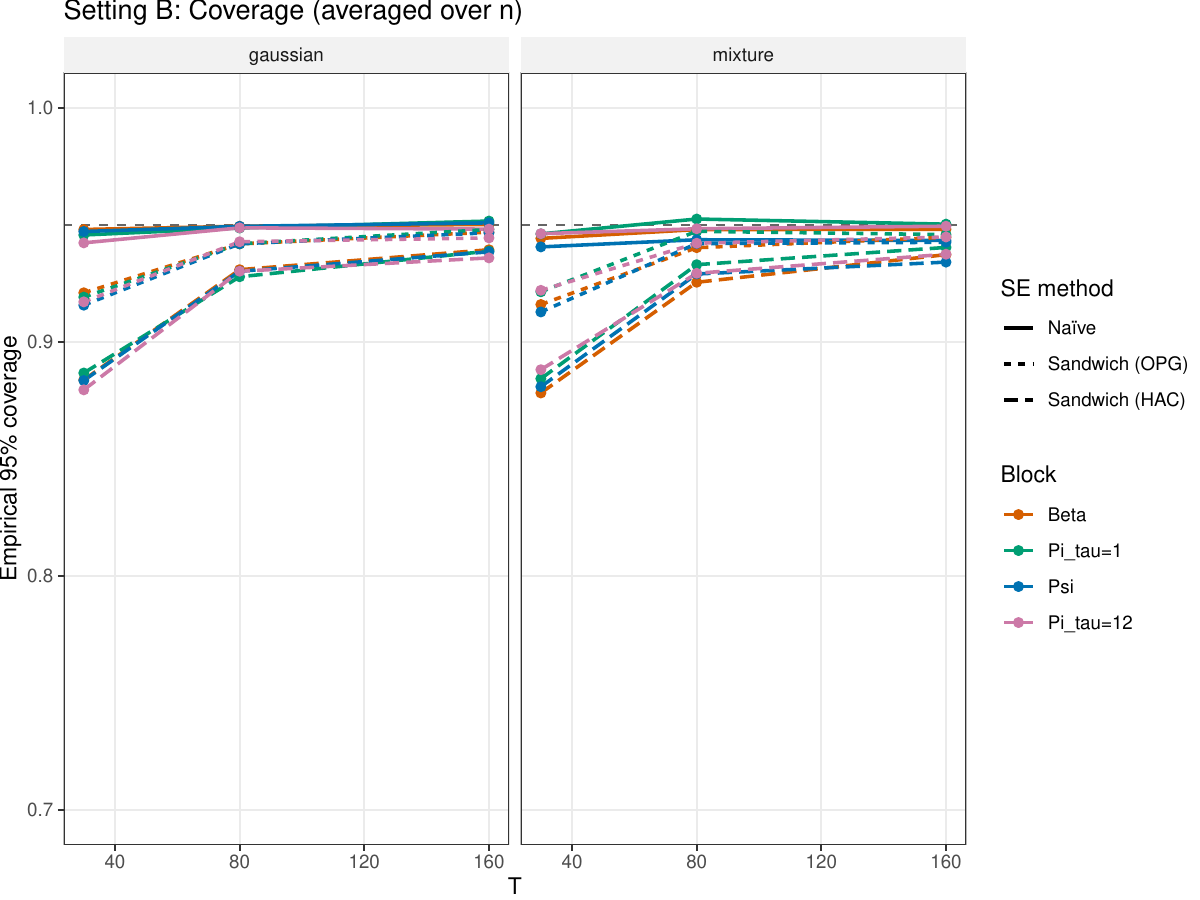}
    \caption{Setting B}
\end{subfigure}
\caption{Finite-sample coverage of naïve, OPG, and HAC variance estimators.}
\label{fig:coverage_sim}
\end{figure}

\textcolor{black}{
Figures~\ref{fig:bias_sim}--\ref{fig:coverage_sim} reveal three broad patterns that are consistent with the dependence regimes induced by Settings~A and~B. First, the average bias (averaged across replications and matrix elements for each component) is small throughout and decreases with $T$ (Figure~\ref{fig:bias_sim}), with the most visible finite-sample bias occurring for the regression block in Setting~B at small $T$. This is expected: the multi-lag temporal component in Setting~B implies substantially higher persistence (spectral radius close to $0.9$), so the effective information accumulation in time is slower and the estimator behaves more like a near-unit-root panel. In contrast, Setting~A (spectral radius around $0.6$) is markedly less persistent, and the estimator is close to unbiased already for moderate $T$. Second, (average) RMSE decreases monotonically in both $n$ and $T$ (Figure~\ref{fig:rmse_sim}). The steep RMSE decline in the regression block is mainly described by improved estimation of the conditional mean as the panel grows, while the relatively flat curves for $\xmat{\Psi}$ and the $\xmat{\Pi}$ blocks reflect that these parameters are already well-identified at the chosen signal-to-noise ratio. The slower convergence in Setting~B is again consistent with high persistence: the process spends longer in correlated ``runs'', reducing the effective sample size in time.}

\textcolor{black}{
The coverage results shown in Figure~\ref{fig:coverage_sim} reflect the different dependence regimes considered in the two simulation settings. In Setting~A, coverage is below the nominal $0.95$ level for small time dimensions, particularly under mixture innovations, but increases steadily with $T$ and approaches the nominal level for larger samples. Differences between the naïve, sandwich (OPG), and HAC variance estimators are generally small in this regime for $T > 30$. In Setting~B, which features stronger dynamic persistence, coverage is similar to setting A, slightly lower though. Robust variance estimators improve coverage in these cases, but the need a sufficient length of the time series. As $T$ increases, coverage stabilises across all parameter blocks and innovation types, and the differences between the three estimators become negligible. Overall, the results indicate that inference improves with the temporal dimension and that robust covariance estimators mainly matter in the more persistent regime and under non-Gaussian innovations.
}

\section{Empirical Results}\label{sec:appl}

In this section, we aim to thoroughly examine the practical implementation of the spatiotemporal autoregressive model as applied to compositional data \textcolor{black}{based on the real-estate market example from above, i.e.,} composition of real estate transactions in Berlin over a 20-year span from 1994 to 2014. \textcolor{black}{In addition to this case study, we discuss the results of a second empirical example in the Appendix, Section \ref{sec:spain}. In contrast to the real-estate transaction compositions, this second case has a much higher spatial resolution, but smaller time horizon.}

\subsection{\textcolor{black}{Parameter estimates in ILR space}}\label{sec:berlin}

\textcolor{black}{Our} empirical investigation examines the spatial and temporal dynamics of real estate transactions within the Berlin metropolitan area, concentrating on three distinct property categories: condominiums, developed plots, and undeveloped plots. This investigation is supported by a dataset containing monthly transaction compositions for all postcode regions in Berlin, covering the period from 1995 to 2015, encompassing $n = 24$ 3-digit postcode areas and $T = 240$ time periods. Through this analysis, we are able to analyse the interactions over time and space, as well as the market share distribution among various property types within the city's internal real estate environment. Such analysis holds particular significance for comprehending the temporal progression of different segments within the urban property market. It illuminates how an increase in the sales of one property type, such as condominiums, might affect the demand for developed or undeveloped plots in surrounding localities. Furthermore, it facilitates the identification of spatial linkages, where patterns in one postcode region may exert influence on neighbouring areas, demonstrating the interconnected nature of the real estate market throughout the city. Figure \ref{fig:berlin1} presents descriptive plots that visualise the spatiotemporal dynamics of the composition data. In the subsequent discussion, extending from the brief introduction in Section \ref{sec:motivation}, we provide additional details for the first case study. Initially, the two left-side plots in the first row demonstrate the compositions for two different years—at the onset and conclusion of the time series—on the simplex. The observation colours correspond to their spatial position, with the map on the right functioning as a colour legend. A subtle spatial dependence is evident, as similar colours tend to cluster nearby on the simplex. Moving to the second row, two simplices illustrate the temporal progression of selected regions, marked on the map by a yellow and a green cross. Both regions exhibit a trend over time shifting towards different property types. Specifically, the green region, represented by postcode area 126xx, transitions from predominantly undeveloped land transactions to a focus on developed plots and condominiums. Finally, the bottom row showcases the time series plots of compositional evolution for these two regions, including a colour reference for the simplices \textcolor{black}{in the middle row}.

Table \ref{tab:results_berlin} shows the estimated coefficients (coefficient matrices) of the spatiotemporal autoregressive process along with \textcolor{black}{robust HAC sandwich standard errors, $t$ statistics, and $p$-values, and summary statistics of the model}. \textcolor{black}{We applied an isometric log-ratio transformation to all compositional observations, using a common orthonormal basis derived from a balanced sequential binary partition that splits the components into groups of approximately equal size.  For the three-part composition, this yields two coordinates: the first contrasts developed land against condominium, while the second contrasts the joint  component (developed land and condominium) against undeveloped land.} The spatial weight matrix has been chosen as a row-standardised contiguity matrix, i.e., all spatial effects should be interpreted as the impact of the average observation of all directly neighbouring regions, sharing common borders. \textcolor{black}{The temporal lags were selected based on AIC and BIC over a set of candidate schemes, which clearly favours the specification $\tau={1,6,12}$ (see Table \ref{tab:model_selection}), indicating that short-run persistence, semi-annual dynamics, and annual seasonality jointly contribute to the temporal dependence structure of the Berlin real-estate market.}

\begin{table}[t]
\centering
\begin{tabular}{l c c c c c}
\hline
Lag set $\tau$ & $\max(\tau)$ & $k$ & Log-likelihood & AIC & BIC \\
\hline
$\{1,6,12\}$  & 12 & 23 & $-10596$ & 21238 & \textbf{21406} \\
$\{1,2,12\}$  & 12 & 23 & $-10612$ & 21270 & 21438 \\
$\{1,3,12\}$  & 12 & 23 & $-10619$ & 21283 & 21451 \\
$\{1,12\}$    & 12 & 19 & $-10950$ & 21937 & 22076 \\
$\{12\}$      & 12 & 15 & $-11824$ & 23677 & 23787 \\
$\{6\}$       & 6  & 15 & $-11965$ & 23961 & 24070 \\
$\{1\}$       & 1  & 15 & $-12151$ & 24331 & 24442 \\
$\{3\}$       & 3  & 15 & $-12181$ & 24392 & 24502 \\
$\{2\}$       & 2  & 15 & $-12190$ & 24410 & 24520 \\
\hline
\end{tabular}
\caption{Model selection for the temporal lag structure based on AIC and BIC. The preferred specification according to both criteria is $\tau=\{1,6,12\}$.}
\label{tab:model_selection}
\end{table}

\begin{table}[t]
\centering
\scriptsize{
\begin{tabular}{cl cccc}
\hline
\multicolumn{2}{l}{Coefficient} & Estimate & Robust SE & $t$ statistic & $p$ value \\
\hline
\multicolumn{6}{l}{\emph{Spatial autoregressive effect $\xmat{\Psi}$}} \\
& $\psi_{1,1}$ & 0.0535 & 0.0183 & 2.9185 & 0.0035 \\
& $\psi_{1,2}$ & 0.0001 & 0.0234 & 0.0037 & 0.9970 \\
& $\psi_{2,1}$ & 0.0529 & 0.0431 & 1.2278 & 0.2195 \\
& $\psi_{2,2}$ & 0.0883 & 0.0330 & 2.6795 & 0.0074 \\
\hline
\multicolumn{6}{l}{\emph{Temporal autoregressive effect (lag 1)}} \\
& $\pi^{(1)}_{1,1}$ & 0.3926 & 0.0165 & 23.8088 & 0.0000 \\
& $\pi^{(1)}_{1,2}$ & 0.0239 & 0.0082 & 2.9058 & 0.0037 \\
& $\pi^{(1)}_{2,1}$ & 0.0786 & 0.0218 & 3.6083 & 0.0003 \\
& $\pi^{(1)}_{2,2}$ & 0.2209 & 0.0158 & 13.9484 & 0.0000 \\
\hline
\multicolumn{6}{l}{\emph{Seasonal autoregressive effect (lag 6)}} \\
& $\pi^{(6)}_{1,1}$ & 0.2804 & 0.0165 & 16.9519 & 0.0000 \\
& $\pi^{(6)}_{1,2}$ & 0.0055 & 0.0080 & 0.6915 & 0.4893 \\
& $\pi^{(6)}_{2,1}$ & -0.0031 & 0.0212 & -0.1442 & 0.8853 \\
& $\pi^{(6)}_{2,2}$ & 0.2370 & 0.0151 & 15.7214 & 0.0000 \\
\hline
\multicolumn{6}{l}{\emph{Seasonal autoregressive effect (lag 12)}} \\
& $\pi^{(12)}_{1,1}$ & 0.1883 & 0.0167 & 11.2663 & 0.0000 \\
& $\pi^{(12)}_{1,2}$ & 0.0094 & 0.0085 & 1.1111 & 0.2665 \\
& $\pi^{(12)}_{2,1}$ & 0.0256 & 0.0231 & 1.1065 & 0.2685 \\
& $\pi^{(12)}_{2,2}$ & 0.2387 & 0.0162 & 14.7499 & 0.0000 \\
\hline
\multicolumn{6}{l}{\emph{Fourier seasonal covariates $\xmat{B}$}} \\
& $\beta_{1,1}$ (intercept coordinate 1) & 0.0508 & 0.0235 & 2.1580 & 0.0309 \\
& $\beta_{1,2}$ (intercept coordinate 2) & -0.0050 & 0.0098 & -0.5172 & 0.6050 \\
& $\beta_{2,1}$ (sine coordinate 1) & 0.0121 & 0.0101 & 1.2016 & 0.2295 \\
& $\beta_{2,2}$ (sine coordinate 2) & 0.1035 & 0.0299 & 3.4649 & 0.0005 \\
& $\beta_{3,1}$ (cosine coordinate 1) & -0.0460 & 0.0151 & -3.0520 & 0.0023 \\
& $\beta_{3,2}$ (cosine coordinate 2) & 0.0211 & 0.0173 & 1.2186 & 0.2230 \\
\hline
\multicolumn{6}{l}{\emph{Innovation variance}} \\
& $\sigma^2$ & 0.4058 & 0.0059 & 68.6253 & 0.0000 \\
\hline
\multicolumn{6}{l}{\emph{Diagnostic checks}} \\
& Residual mean (overall) & $1.780 \times 10^{-6}$ &  &  &  \\[.1cm]
& Ljung--Box (pooled residuals) & 17.79  & ($p=0.1222$) &  &  \\
& ACF(1) average, $k=1$ (Prop. rejected $\alpha=0.05$ LB) & -0.167 & ($0.7917$) &  &  \\
& ACF(1) average, $k=2$ (Prop. rejected $\alpha=0.05$ LB) & -0.121 & ($0.7083$) &  &  \\
& Moran's $I$ average, $k=1$ (Prop. rejected $\alpha=0.05$) & $-0.043$ & ($0.0614$) &  &  \\
& Moran's $I$ average, $k=2$ (Prop. rejected $\alpha=0.05$) & $-0.053$ & ($0.0394$) &  &  \\
& Excess kurtosis (pooled) & 1.295 &  &  &  \\
& Spectral radius $<1$ & 0.7411 &  &  &  \\
\hline
\multicolumn{6}{l}{\emph{Goodness of fit}} \\
& Coefficient of determination (ILR $k=1$) & 0.639 &  &  &  \\
& Coefficient of determination (ILR $k=2$) & 0.362 &  &  &  \\
& RMSE (Developed land) & 0.083 &  &  &  \\
& RMSE (Condominium) & 0.122 &  &  &  \\
& RMSE (Undeveloped land) & 0.095 &  &  &  \\
& Aitchison RMSE & 0.755 &  &  &  \\
\hline
\end{tabular}
}
\caption{Estimated parameters of the extended spatiotemporal multivariate autoregressive model for the Berlin real-estate data. Robust standard errors are computed using a HAC sandwich estimator based on the long-run variance.}
\label{tab:results_berlin}
\end{table}

\textcolor{black}{Upon inspection of the estimated parameters in Table~\ref{tab:results_berlin}, temporal dependence dominates contemporaneous spatial interaction. All diagonal elements of the lag-1 temporal matrix $\xmat{\Pi}_{1}$ are large and statistically significant, indicating strong short-run persistence in both ilr coordinates. In addition, the diagonal elements of the seasonal lag-12 matrix $\xmat{\Pi}_{12}$ are also significant, providing clear evidence of annual recurrence in the compositional dynamics. The off-diagonal elements of $\xmat{\Pi}_{1}$ are statistically significant but substantially smaller in magnitude than the diagonal terms, indicating modest cross-coordinate temporal transmission. In contrast, off-diagonal elements of the higher-order temporal lags are not statistically significant, suggesting that seasonal feedback operates primarily within, rather than across, ilr coordinates. Spatial autoregressive effects are comparatively weak. Under robust HAC inference, the diagonal elements $\psi_{1,1}=0.0535$ and $\psi_{2,2}=0.0883$ is statistically significant at an $\alpha=0.05$ level, while the off-diagonal entries are not significant. This indicates that contemporaneous spatial dependence is present but limited, and concentrated in the second ilr balance, rather than uniformly across both coordinates.}

\textcolor{black}{Residual diagnostics indicate that the model adequately captures the main spatiotemporal dependence structure. The pooled Ljung–Box test does not reject global serial independence, and average residual autocorrelations is significantly different from zero for about two third of the locations, while Moran’s I statistics are close to zero, suggesting no systematic remaining spatial autocorrelation. The negative autocorrelation of the residuals indicates a slight overcompensation of the temporal autocorrelation. Although residuals exhibit moderate excess kurtosis and deviations from Gaussianity, stability conditions are satisfied and robust HAC inference accounts for remaining weak dependence and distributional departures. The reported goodness-of-fit measures indicate that the model captures a substantial share of the variation in the first ilr coordinate and a moderate share in the second, while the root mean squared errors on the simplex suggest a reasonably accurate fit across all three market segments, with the smallest prediction error observed for developed land and slightly larger deviations for condominiums.}

\subsection{\textcolor{black}{Economic interpretation in original compositional space}}\label{sec:berlin2}

\begin{table}[t]
\centering
\textcolor{black}{
\begin{tabular}{l r r r}
\hline
 & Direct & Indirect (average) & Total (sum) \\
\hline
\multicolumn{4}{l}{\emph{Average baseline intercept effect}}\\
\multicolumn{4}{l}{\emph{(permanent level shift in the baseline seasonal trajectory)}}\\
$\quad$ Developed land   & $-0.0090$ & $-0.0000$ & $-0.0093$ \\
$\quad$ Condominium      & $ 0.0221$ & $ 0.0001$ & $ 0.0240$ \\
$\quad$ Undeveloped land & $-0.0132$ & $-0.0001$ & $-0.0147$ \\
\hline
\multicolumn{4}{l}{\emph{Location-specific effect at unit $j=24$: unit innovation in ilr coordinate $k=1$}}\\
\multicolumn{4}{l}{\emph{(first coordinate; propagated via spatial multiplier)}}\\
$\quad$ Developed land   & $-0.2355$ & $-0.0006$ & $-0.2499$ \\
$\quad$ Condominium      & $ 0.2668$ & $ 0.0011$ & $ 0.2920$ \\
$\quad$ Undeveloped land & $-0.0313$ & $-0.0005$ & $-0.0420$ \\
\hline
\multicolumn{4}{l}{\emph{Location-specific effect at unit $j=24$: unit innovation in ilr coordinate $k=2$}}\\
\multicolumn{4}{l}{\emph{(second coordinate; propagated via spatial multiplier)}}\\
$\quad$ Developed land   & $ 0.0290$ & $ 0.0002$ & $ 0.0324$ \\
$\quad$ Condominium      & $ 0.0831$ & $ 0.0004$ & $ 0.0931$ \\
$\quad$ Undeveloped land & $-0.1121$ & $-0.0006$ & $-0.1256$ \\
\hline
\end{tabular}
}
\caption{Simplex-scale effects evaluated at the reference composition. The first block reports average direct, indirect, and total effects of a one-unit increase in the Fourier intercept term, corresponding to a permanent shift in the baseline seasonal trajectory. The second and third blocks report location-specific effects of a unit innovation shock at spatial unit $j=24$ (postcode area 126xx) in the first and second ilr coordinates, respectively, propagated through the spatial multiplier. Direct effects denote own-unit impacts, indirect effects denote average spillovers to the remaining locations, and total effects combine direct and spatial spillover contributions across all locations. Components are ordered as developed land, condominium, and undeveloped land.}
\label{tab:baseline_spatial_effects}
\end{table}

\textcolor{black}{Interpreting these findings on the original compositional scale requires caution due to the non-linear nature of the inverse ilr transformation. All reported effects on the simplex are evaluated locally at a representative reference composition, chosen as the overall mean composition across space and time. This choice reflects the typical market structure in the sample and ensures that reported effects correspond to empirically relevant states of the system. Since the covariate specification is seasonal (Fourier terms), the intercept component represents the baseline level of the seasonal trajectory in ilr space around which periodic fluctuations occur. When mapped back to the simplex and evaluated at the mean reference composition, the implied effects describe systematic reallocations of sales shares across segments rather than uniform shifts. As reported in Table~\ref{tab:baseline_spatial_effects}, the baseline intercept effect primarily operates through direct (own-location) adjustments, while spatial spillovers are economically negligible on average. Quantitatively, the baseline shift increases the share of \emph{condominium} transactions by about $2.2$ percentage points and reduces the shares of \emph{developed land} and \emph{undeveloped land} by roughly $0.9$ and $1.3$ percentage points, respectively. This pattern suggests a gradual reallocation of market activity towards condominium transactions within the urban housing market, while the relative importance of land transactions declines. Such a shift is consistent with mature metropolitan markets, where development activity increasingly concentrates on condominium and apartment segments rather than on land turnover.}

\textcolor{black}{To illustrate contemporaneous spatial propagation, Table~\ref{tab:baseline_spatial_effects} also reports the immediate simplex-scale effects of unit innovation shocks in both ilr coordinates at a representative district ($j=24$, postcode area 126xx, green compositions in Figure \ref{fig:berlin1}). By construction of the ilr basis, the first coordinate contrasts \emph{developed land} with \emph{condominium}, while the second coordinate contrasts the joint component \emph{(developed land and condominium)} against \emph{undeveloped land}. For a unit innovation in the first coordinate ($k=1$), the direct effect at the shocked district implies a substantial reallocation from \emph{developed land} towards \emph{condominium} transactions. At the reference composition, the condominium share increases by roughly $26.7$ percentage points, while developed land decreases by about $23.6$ percentage points and undeveloped land declines slightly (around $3.1$ percentage points). Spatial spillovers transmitted through the spatial multiplier remain small but reinforce the same directional adjustment, yielding total effects of approximately $+29.2$ percentage points for condominiums, $-25.0$ percentage points for developed land, and $-4.2$ percentage points for undeveloped land. A unit innovation in the second coordinate ($k=2$) generates a different compositional adjustment: the shares of \emph{developed land} and \emph{condominium} transactions increase moderately (about $2.9$ and $8.3$ percentage points, respectively), while the share of \emph{undeveloped land} decreases by roughly $11.2$ percentage points. Again, spatial spillovers remain small but slightly amplify the response, resulting in total effects of approximately $+3.2$, $+9.3$, and $-12.6$ percentage points for developed land, condominiums, and undeveloped land, respectively. Economically, this balance represents a shift of market activity away from undeveloped land and towards developed real estate segments. Taken together, these results indicate that local shocks primarily affect the composition of transactions within the originating district, while spatial propagation plays a secondary role. The dominant adjustment occurs through substitution between developed market segments and reductions in the relative share of undeveloped land transactions, which is consistent with the highly urbanised structure of the Berlin housing market.}

\begin{figure}[t]
\centering
\includegraphics[width=0.48\textwidth]{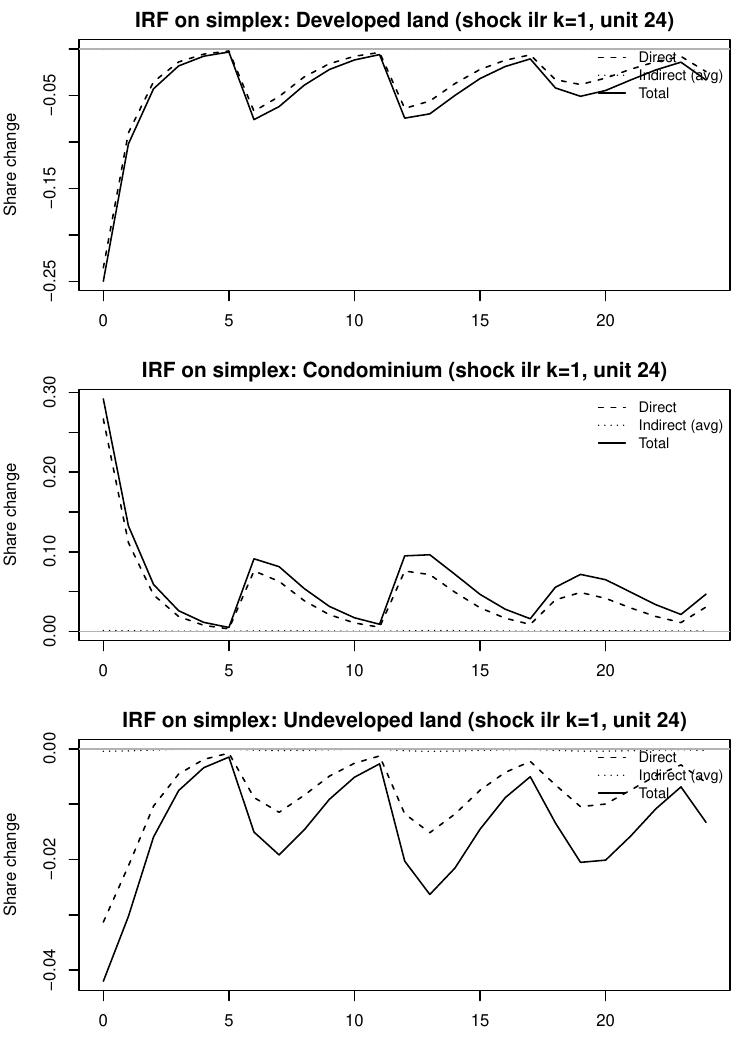}
\includegraphics[width=0.48\textwidth]{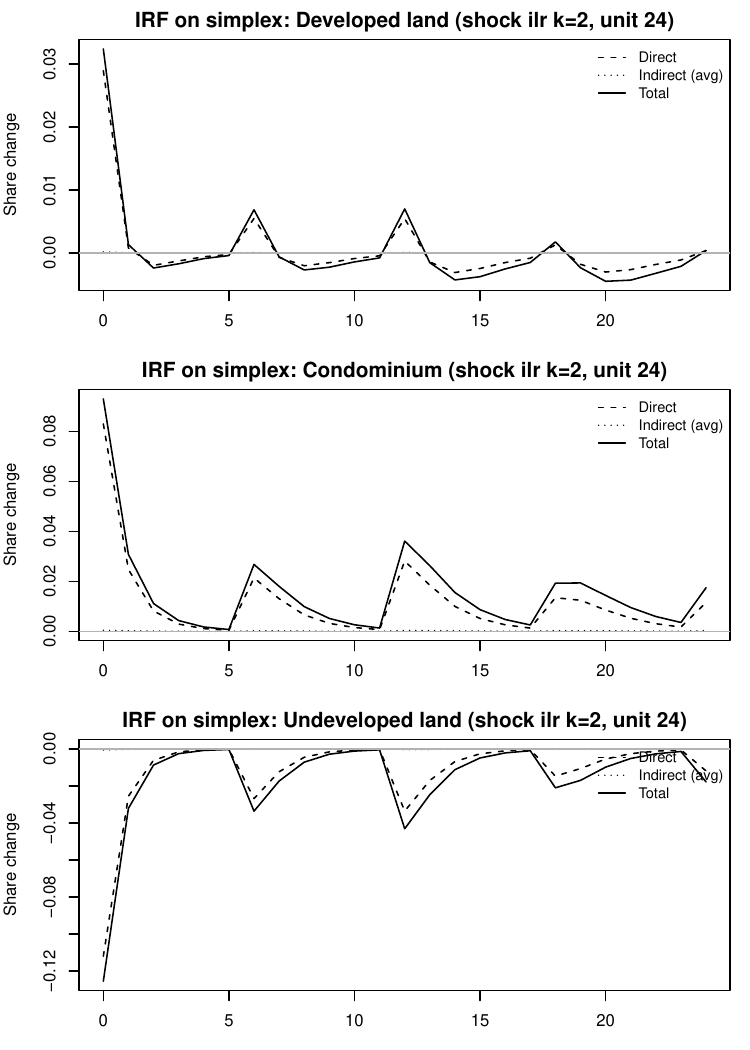}
\caption{Impulse response functions on the simplex for a unit innovation in ilr coordinate $k=1$ (left) and $k = 2$ (right) at a single spatial unit. The figure reports direct effects (own-unit), indirect effects (average spillovers), and total effects for each compositional component over time.}
\label{fig:irf_simplex}
\end{figure}

\textcolor{black}{Temporal dependence further shapes the evolution of compositional adjustments. Figure~\ref{fig:irf_simplex} presents impulse response functions on the simplex for unit innovations in ilr coordinates $k=1$ (left column) and $k=2$ (right column) at the same spatial unit ($j=24$, eastern part, postcode area 126xx). All responses incorporate both spatial and temporal feedback. The impulse responses remain bounded and gradually decay over time, confirming that the estimated spatiotemporal system is dynamically stable. The immediate effects differ across the two balances implied by the ilr coordinates. A shock in the first coordinate ($k=1$) primarily reallocates activity from \emph{developed land} towards \emph{condominium} transactions, while the share of \emph{undeveloped land} declines moderately. In contrast, a shock in the second coordinate ($k=2$) generates a stronger shift away from \emph{undeveloped land}, accompanied by moderate increases in both \emph{developed land} and \emph{condominium} transactions.  Following the initial response, the impulse responses decay over the first few periods, reflecting strong short-run persistence driven by the first-order temporal lag. At longer horizons, smaller secondary peaks appear around the 6- and 12-month horizons, indicating seasonal feedback effects captured by the higher-order lag matrices $\xmat{\Pi}_{6}$ and $\xmat{\Pi}_{12}$. These recurring but damped responses suggest that part of the initial disturbance re-enters the system through seasonal transaction cycles. Across all horizons, direct (own-unit) effects account for the bulk of the response, while average spillover effects across neighbouring districts remain comparatively small. Consequently, spatial interactions mainly propagate the shock locally without substantially altering the overall dynamic pattern. Taken together, the impulse responses indicate that shocks in the Berlin housing market primarily trigger within-district compositional reallocation between developed market segments and undeveloped land, while seasonal persistence and modest spatial feedback govern the subsequent adjustment path.}

\section{Conclusion}\label{sec:conclusion}

Understanding how regional economic structures evolve over space and time requires models that accommodate both spatial and temporal dependence while respecting the compositional nature of the data. In this paper, we addressed this challenge through \textcolor{black}{a} real-world applications: the monthly evolution of housing market compositions across postcode areas in Berlin. \textcolor{black}{The example highlighted} key features of compositional areal data in practice---structural constraints, spatiotemporal dependence, and component-specific dynamics, which conventional models would fail to accommodate.

To address these challenges, we proposed a spatiotemporal multivariate autoregressive framework tailored for composition-valued panel data. The model combines a quasi maximum likelihood estimator with isometric log-ratio transformations, ensuring identifiability and asymptotic consistency under increasing space and time domains. In the Berlin housing market, we found evidence of moderate spatial spillovers and strong persistence in the share of undeveloped land. Cross-component interactions were weak in both cases, but statistically significant.

The flexibility and interpretability of the model make it a valuable tool for analysing compositionally constrained processes in regional economics and related fields. Future research could extend the framework in several directions. Methodologically, non-Gaussian extensions, such as compositional count models or models for zero-inflated data, could allow applications in domains like ecology, demography, or epidemiology. Allowing spatially or temporally varying autoregressive coefficients may help capture localised dynamics or structural heterogeneity. Furthermore, models that incorporate covariates in both the mean and dependency structure, or that allow for endogenous lagged effects across compositional components, would increase applicability. Finally, change-point extensions or regime-switching versions of the model could account for structural breaks or policy-induced shifts in regional dynamics.




\newpage


\section*{Appendix}

\subsection*{Second empirical example: Spanish economic sectors}\label{sec:spain}

The second dataset is derived from information contained within the Spanish Central Business Register, and was obtained from the official website (\url{www.ine.es}) of the National Statistics Institute of Spain (INE). For this study, we analyse regional data on local business compositions, segmented into three categories, collected annually for 2793 administrative regions \textcolor{black}{between 2012 and 2021. Specifically,} our concentrated examination addressed several key sectors, notably: \textbf{services}, which include a broad spectrum of activities such as communication services, financial and insurance services, administrative and support services, educational services, health and social care services, in addition to artistic, recreational, and entertainment pursuits; the \textbf{industry} sector, which encompasses a range of activities from extractive and manufacturing industries to energy supply, water supply, sanitation services, waste management, and contamination remediation activities; and the \textbf{construction} sector.

The records provide the total number of firms operating in each sector within every municipality. To avoid inconsistencies caused by firms operating in multiple locations, INE assigns each company to a single location unit based on the registered address of the corporate headquarters. The resulting sector counts were subsequently transformed into compositions using the closure operator. With this data as input, all absolute figures were converted into local business sector compositions by employing the closure operator in a subsequent process.

\begin{figure}
    \centering
    \includegraphics[width=0.32\linewidth]{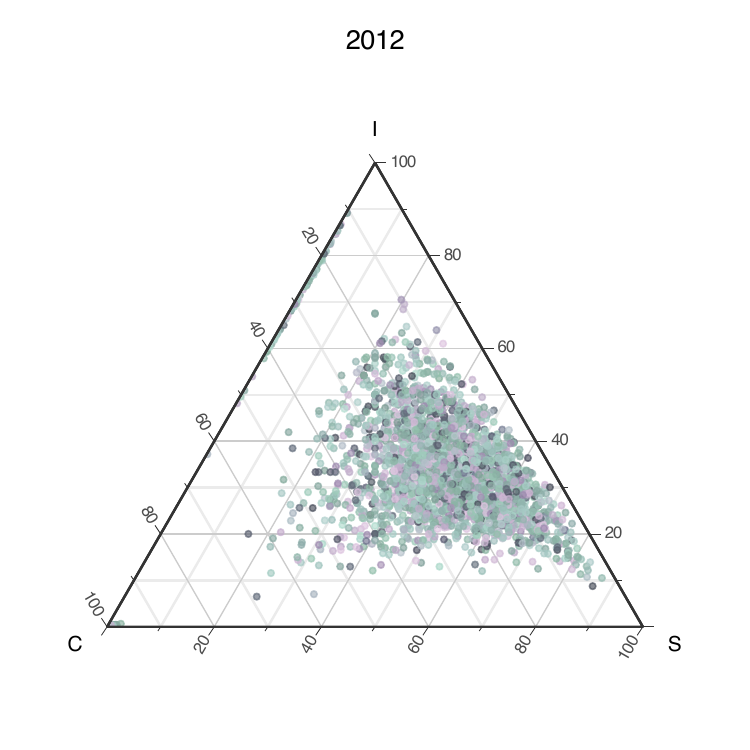}
    \includegraphics[width=0.32\linewidth]{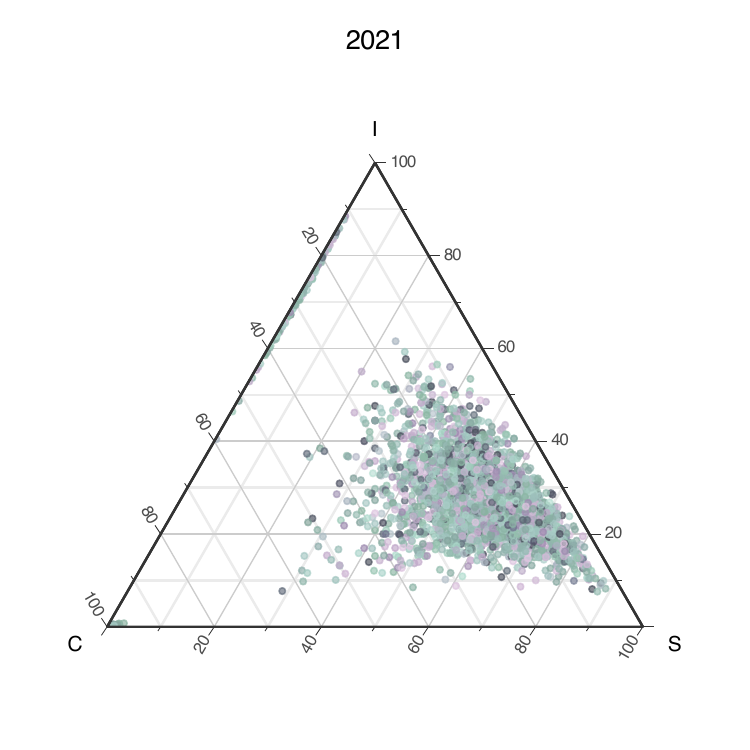} 
    \includegraphics[width=0.32\linewidth]{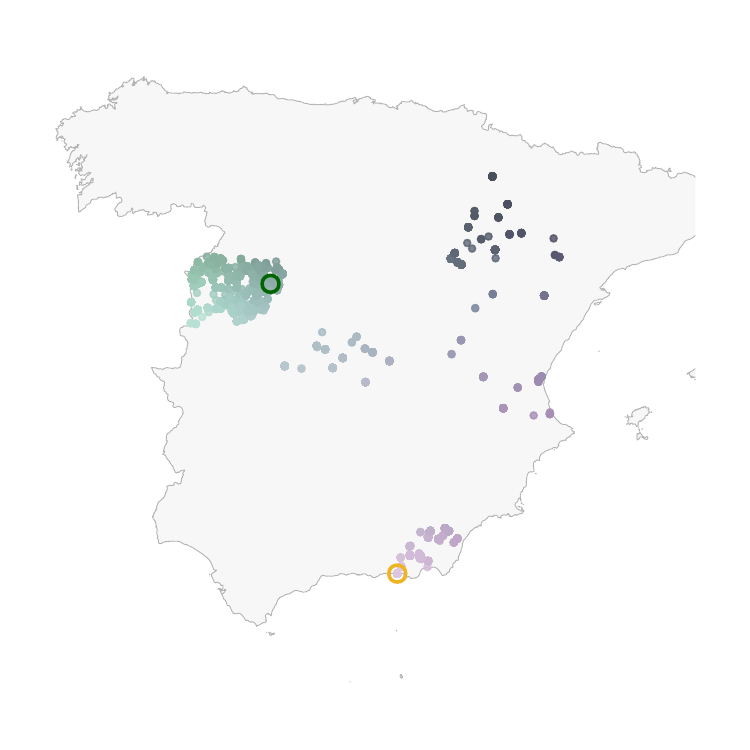} \\
    \includegraphics[width=0.43\linewidth]{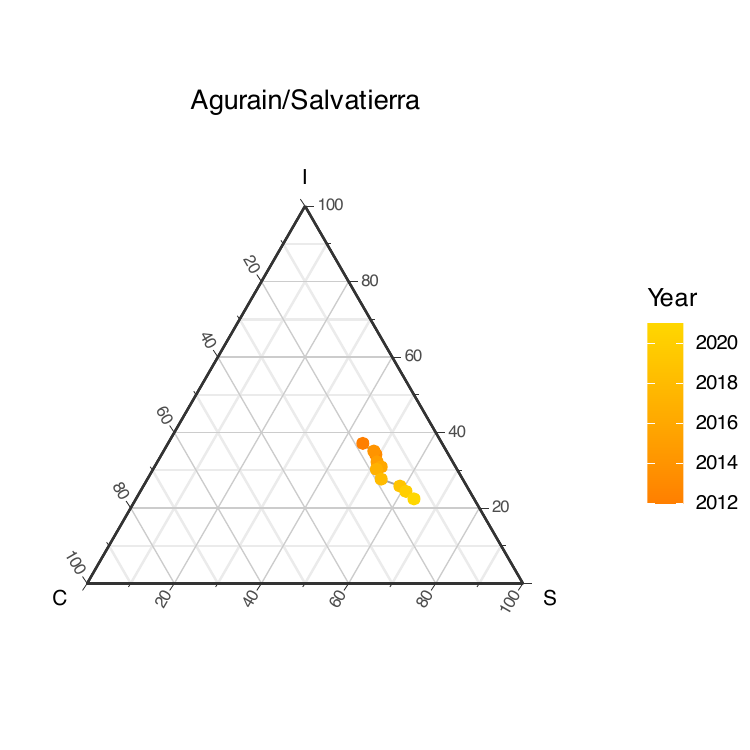}
    \includegraphics[width=0.43\linewidth]{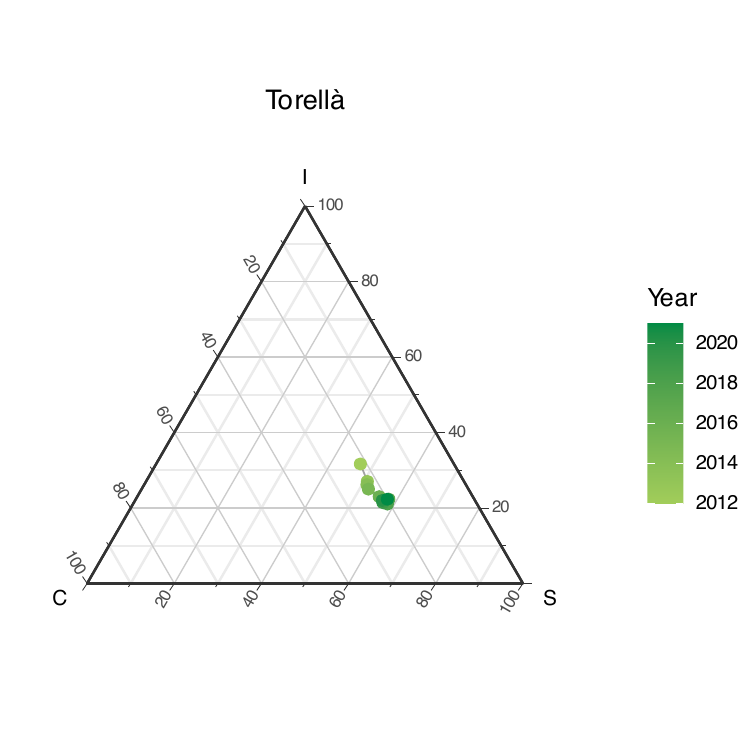}\\
    \includegraphics[width=0.43\linewidth]{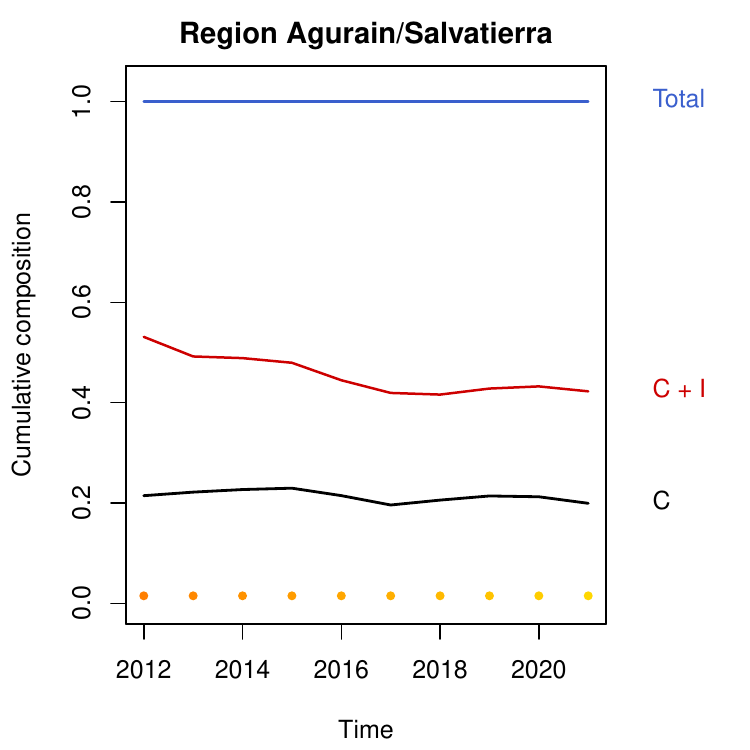}
    \includegraphics[width=0.43\linewidth]{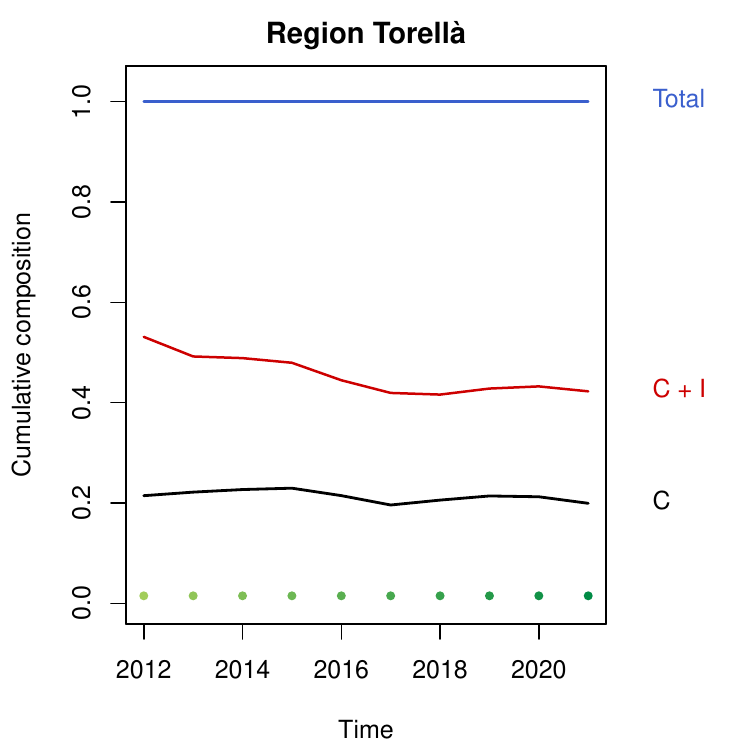}
    \caption{Overview of the Spanish economic sector composition data set. First row: Compositions of  (I: Industry, C: Construction, S: Services) in 2012 (left) and 2021 (centre). The points are coloured according to the spatial location as shown in the map legend (right). Middle row: Composition of two selected regions, where the points are coloured according to the time points. The colours correspond to the colour of the marks on the map. Bottom row: Compositions of the same two locations displayed as time series plots.}
    \label{fig:spain1}
\end{figure}

\textcolor{black}{Analogous to the Berlin example, descriptive plots for the Spanish data are shown in Figure~\ref{fig:spain1}.} Unlike the Berlin example, there is no apparent spatial dependency observed within the compositions. The colours in the plots on the first row do not reveal discernible patterns. Nevertheless, a distinct temporal pattern emerges within the data, though the time frame of the series is notably shorter. Notably, in Agurain/Salvatierra (highlighted in yellow), a downward trend in the construction sector is observed, benefiting the service sector, whereas the industrial sector's share remained consistently high over time.

These empirical observations are largely supported by the estimates of the spatiotemporal autoregressive model reported in Table~\ref{tab:results_spain}. The isometric log-ratio (ilr) transformation was applied in the same manner as in the Berlin example. Because the municipalities are irregularly distributed across Spain, \textcolor{black}{the spatial dependence structure was defined using a row-standardised distance-based weight matrix. Specifically, locations within a fixed distance threshold were assigned equal weights, resulting in a sparse spatial weight matrix with approximately 94\% zero entries. Sensitivity analyses with alternative distance cut-offs yielded very similar estimation results, indicating that the findings are robust with respect to the spatial specification.}

\textcolor{black}{The estimated parameters highlight the dominance of temporal dependence in the evolution of the sectoral compositions. The diagonal elements of the temporal autoregressive matrix $\hat{\xmat{\Pi}}$ are extremely large and highly significant, with $\pi_{1,1}=0.967$ and $\pi_{2,2}=0.951$. These values indicate very strong persistence in both ilr coordinates, implying that the sectoral compositions evolve slowly over time and remain close to their previous states. The off-diagonal elements are considerably smaller. While $\pi_{2,1}=-0.020$ is statistically significant, its magnitude is negligible relative to the diagonal coefficients, and $\pi_{1,2}=-0.020$ is only weakly significant at the 10\% level. Overall, cross-coordinate interactions appear to play only a minor role in the temporal dynamics.}

\textcolor{black}{In contrast to the Berlin case, the estimated spatial autoregressive effects are very small and statistically insignificant. All entries of the spatial coefficient matrix $\hat{\xmat{\Psi}}$ are close to zero, with estimates ranging from $0.000$ to $0.018$. None of these effects are statistically significant at conventional levels, indicating that spatial spillovers between neighbouring municipalities are negligible in this dataset. This finding is consistent with the descriptive maps, which showed no clear spatial clustering of sectoral compositions.}

\textcolor{black}{The diagnostic measures further support the adequacy of the model. The coefficients of determination for the two ilr coordinates are very high, with $R^2_1=0.931$ and $R^2_2=0.917$, indicating that the model explains most of the temporal variation in the transformed compositions. The prediction errors on the simplex scale are also small, with RMSE values of approximately $0.051$ for construction, $0.042$ for industry, and $0.045$ for services. The average Aitchison RMSE equals $0.157$.}

\begin{table}[t]
\centering
\scriptsize{
\begin{tabular}{cl cccc}
\hline
\multicolumn{2}{l}{Coefficient} & Estimate & Robust SE & $t$ statistic & $p$ value \\
\hline
\multicolumn{6}{l}{\emph{Spatial autoregressive effect $\xmat{\Psi}$}} \\
& $\psi_{1,1}$ & 0.0000 & 0.0365 & 0.0000 & 1.0000 \\
& $\psi_{1,2}$ & 0.0042 & 0.0197 & 0.2130 & 0.8313 \\
& $\psi_{2,1}$ & 0.0181 & 0.0194 & 0.9349 & 0.3499 \\
& $\psi_{2,2}$ & 0.0115 & 0.0164 & 0.6999 & 0.4840 \\
\hline
\multicolumn{6}{l}{\emph{Temporal autoregressive effect (lag 1)}} \\
& $\pi^{(1)}_{1,1}$ & 0.9673 & 0.0051 & 191.2605 & 0.0000 \\
& $\pi^{(1)}_{1,2}$ & -0.0200 & 0.0105 & -1.9023 & 0.0571 \\
& $\pi^{(1)}_{2,1}$ & -0.0198 & 0.0029 & -6.8951 & 0.0000 \\
& $\pi^{(1)}_{2,2}$ & 0.9507 & 0.0057 & 166.2886 & 0.0000 \\
\hline
\multicolumn{6}{l}{\emph{Intercept $\xmat{B}$}} \\
& $\beta_{1,1}$ (intercept coordinate 1) & -0.0198 & 0.0073 & -2.7164 & 0.0066 \\
& $\beta_{1,2}$ (intercept coordinate 2) & 0.0259 & 0.0066 & 3.9166 & 0.0001 \\
\hline
\multicolumn{6}{l}{\emph{Innovation variance}} \\
& $\sigma^2$ & 0.0617 & 0.0049 & 12.7275 & 0.0000 \\
\hline
\multicolumn{6}{l}{\emph{Diagnostic checks}} \\
& Residual mean (overall) & $5.025 \times 10^{-8}$ &  &  &  \\[.1cm]
& Ljung--Box (pooled residuals) & 0.0255 & ($p=0.8731$) &  &  \\
& ACF(1) average, $k=1$ (Prop. rejected $\alpha=0.05$ LB) & -0.1152 & ($0.0816$) &  &  \\
& ACF(1) average, $k=2$ (Prop. rejected $\alpha=0.05$ LB) & -0.1596 & ($0.0795$) &  &  \\
& Moran's $I$ average, $k=1$ (Prop. rejected $\alpha=0.05$) & -0.0014 & ($0$) &  &  \\
& Moran's $I$ average, $k=2$ (Prop. rejected $\alpha=0.05$) & -0.0005 & ($0$) &  &  \\
& Excess kurtosis (pooled) & 190.844 &  &  &  \\
\hline
\multicolumn{6}{l}{\emph{Goodness of fit}} \\
& Coefficient of determination (ILR $k=1$) & 0.931 &  &  &  \\
& Coefficient of determination (ILR $k=2$) & 0.917 &  &  &  \\
& RMSE (Construction) & 0.051 &  &  &  \\
& RMSE (Industry) & 0.042 &  &  &  \\
& RMSE (Services) & 0.045 &  &  &  \\
& Aitchison RMSE & 0.157 &  &  &  \\
\hline
\end{tabular}
}
\caption{Estimated parameters of the spatiotemporal multivariate autoregressive model for the Spanish regional business composition data (2012--2021). Robust standard errors are computed using a sandwich estimator based on the long-run variance.}
\label{tab:results_spain}
\end{table}

\newpage

\subsection*{Proofs of the theorems}

The following supplementary material provides the proofs of the theoretical results.

\begin{proof}[Proof of Theorem \ref{th:consistency}]

The proof of the theorem consists of two main steps. First, the identification of the parameters is established by showing that the population objective is uniquely maximised at $(\vartheta_0,\sigma_0^2)$. Second, we prove uniform convergence of the
scaled sample criterion to its population counterpart. Finally, consistency follows from an argmax theorem.

\medskip
Let $m = n(D-1)$ and abbreviate $\xmat{S}(\xmat{\Psi})=\xmat{S}_{n(D-1)}(\xmat{\Psi})$. Define the scaled sample criterion (up to constants independent of $(\vartheta,\sigma^2)$)
\begin{equation}\label{eq:ellnT_pf}
\ell_{nT}(\vartheta,\sigma^2)
=
\frac{1}{mT}\log\mathcal{L}_{nT}(\vartheta,\sigma^2)
=
\frac{1}{m}\log|\xmat{S}(\xmat{\Psi})|
-\frac{1}{2}\log\sigma^2
-\frac{1}{2\sigma^2}\cdot\frac{1}{mT}\sum_{t=1}^T \xvec{r}_t(\vartheta)^\top \xvec{r}_t(\vartheta)
+c_{nT},
\end{equation}
where $c_{nT}$ is constant in $(\vartheta,\sigma^2)$. Let $Q(\vartheta,\sigma^2)=\E[\ell_{nT}(\vartheta,\sigma^2)]$ denote the population objective.

\medskip
\noindent\textbf{Step 1: Identification.}
We show that, for all $(\vartheta,\sigma^2)\in\Theta\times[\underline\sigma^2,\overline\sigma^2]$,
\begin{equation}\label{eq:ident_pf}
Q(\vartheta,\sigma^2)-Q(\vartheta_0,\sigma_0^2)\le 0,
\end{equation}
with equality if and only if $(\vartheta,\sigma^2)=(\vartheta_0,\sigma_0^2)$.

Let $\xmat{S}_0=\xmat{S}(\xmat{\Psi}_0)$ and, for a finite set of temporal lags 
$\{\tau_1,\dots,\tau_L\}$, define
\[
\xmat{A}_{0,\ell}
=
\xmat{A}(\xmat{\Pi}_{0,\ell})
=
\xmat{\Pi}_{0,\ell}^{\top}\otimes \xmat{I}_n,
\qquad \ell=1,\dots,L.
\]
Since the set of lags $\{\tau_1,\dots,\tau_L\}$ is finite and fixed, the process admits a finite-dimensional companion representation. From the reduced form under the DGP,
\begin{equation}\label{eq:DGP_pf}
\ddot{\xvec{Y}}_t
=
\xmat{S}_0^{-1}
\Big(
\xvec{x}_t(\xmat{B}_0)
+
\sum_{\ell=1}^{L}
\xmat{A}_{0,\ell}\ddot{\xvec{Y}}_{t-\tau_\ell}
+
\xvec{u}_t
\Big).
\end{equation}
Define
\[
\xmat{B}(\xmat{\Psi})
=
\xmat{S}(\xmat{\Psi})\,\xmat{S}_0^{-1},
\qquad
\xmat{M}(\xmat{\Psi})
=
\xmat{B}(\xmat{\Psi})^\top \xmat{B}(\xmat{\Psi})\succ 0.
\]
Using \eqref{eq:DGP_pf}, for any $\vartheta=(\xmat{B},\xmat{\Psi},\xmat{\Pi})\in\Theta$ the residuals admit the decomposition
\begin{align}
\xvec{r}_t(\vartheta)
=
\xmat{S}(\xmat{\Psi})\,\ddot{\xvec{Y}}_t
-\xvec{x}_t(\xmat{B})
-
\sum_{\ell=1}^{L}
\xmat{A}_{\ell}(\xmat{\Pi}_{\ell})
\ddot{\xvec{Y}}_{t-\tau_\ell},\label{eq:r_decomp_pf}
\end{align}
where
\begin{equation}\label{eq:Delta_pf}
\Delta_t(\vartheta)
=
\xmat{B}(\xmat{\Psi})
\Big(
\xvec{x}_t(\xmat{B}_0)
+
\sum_{\ell=1}^{L}
\xmat{A}_{0,\ell}\ddot{\xvec{Y}}_{t-\tau_\ell}
\Big)
-
\xvec{x}_t(\xmat{B})
-
\sum_{\ell=1}^{L}
\xmat{A}_{\ell}(\xmat{\Pi}_{\ell})
\ddot{\xvec{Y}}_{t-\tau_\ell}.
\end{equation}
Let $\mathcal{F}_{t-1}=\sigma\{\xvec{u}_s,\xmat{X}_s:s\le t-1\}$ and $\mathcal{G}_t=\sigma(\mathcal{F}_{t-1},\xmat{X}_t)$.
Then $\Delta_t(\vartheta)$ is $\mathcal{G}_t$-measurable and, by Assumption~\ref{ass:err_exo_cons},
$\E(\xvec{u}_t\mid\mathcal{G}_t)=\xvec{0}$. Hence,
\[
\E\!\left[\xvec{u}_t^\top \xmat{B}(\xmat{\Psi})^\top \Delta_t(\vartheta)\right]=0,
\]
and therefore
\begin{equation}\label{eq:Ert_pf}
\E\!\left[\xvec{r}_t(\vartheta)^\top\xvec{r}_t(\vartheta)\right]
=
\E\!\left[\xvec{u}_t^\top \xmat{M}(\xmat{\Psi})\,\xvec{u}_t\right]
+\E\!\left[\Delta_t(\vartheta)^\top\Delta_t(\vartheta)\right]
=
\sigma_0^2\,\tr\!\big(\xmat{M}(\xmat{\Psi})\big)
+\E\!\left[\Delta_t(\vartheta)^\top\Delta_t(\vartheta)\right],
\end{equation}
where we used $\Var(\xvec{u}_t)=\sigma_0^2\xmat{I}_m$ from Assumption~\ref{ass:err_exo_cons}.

For any $m\times m$ matrix $\xmat{M}\succ0$,
\begin{equation}\label{eq:trlog_pf}
\tr(\xmat{M})-\log|\xmat{M}|-m\ge 0,
\qquad\text{with equality iff }\xmat{M}=\xmat{I}_m.
\end{equation}
Since $\xmat{M}(\xmat{\Psi})=\xmat{B}(\xmat{\Psi})^\top\xmat{B}(\xmat{\Psi})$ and
$\log|\xmat{M}(\xmat{\Psi})|
=2\log|\xmat{B}(\xmat{\Psi})|
=2\big(\log|\xmat{S}(\xmat{\Psi})|-\log|\xmat{S}_0|\big)$,
\eqref{eq:trlog_pf} implies
\begin{equation}\label{eq:det_trace_link_pf}
\frac{1}{m}\big(\log|\xmat{S}(\xmat{\Psi})|-\log|\xmat{S}_0|\big)
\le
\frac{1}{2m}\Big(\tr\big(\xmat{M}(\xmat{\Psi})\big)-m\Big),
\end{equation}
with equality if and only if $\xmat{M}(\xmat{\Psi})=\xmat{I}_m$, i.e.\ $\xmat{S}(\xmat{\Psi})=\xmat{S}_0$.
Under Assumption~\ref{ass:stability_cons} and the affine form
$\xmat{S}(\xmat{\Psi})=\xmat{I}_m-\xmat{\Psi}^\top\otimes\xmat{W}_n$, the identity
$\xmat{S}(\xmat{\Psi})=\xmat{S}_0$ implies $\xmat{\Psi}=\xmat{\Psi}_0$.\footnote{This holds provided $\xmat{W}_n\neq \xmat{0}$.}

Using \eqref{eq:ellnT_pf} and \eqref{eq:Ert_pf} and the fact that $\E(\xvec{u}_t^\top\xvec{u}_t)=m\sigma_0^2$, we obtain
\begin{align*}
Q(\vartheta,\sigma^2)-Q(\vartheta_0,\sigma_0^2)
&=
\frac{1}{m}\big(\log|\xmat{S}(\xmat{\Psi})|-\log|\xmat{S}_0|\big)
-\frac{1}{2}\log\frac{\sigma^2}{\sigma_0^2}\\
&\quad
-\frac{1}{2\sigma^2}\cdot\frac{1}{m}\Big(\sigma_0^2\tr(\xmat{M}(\xmat{\Psi}))
+\E[\Delta_t(\vartheta)^\top\Delta_t(\vartheta)]\Big)
+\frac{1}{2}.
\end{align*}
Dropping the non-positive term $-\frac{1}{2\sigma^2 m}\E[\Delta_t(\vartheta)^\top\Delta_t(\vartheta)]$ and applying
\eqref{eq:det_trace_link_pf} yields
\[
Q(\vartheta,\sigma^2)-Q(\vartheta_0,\sigma_0^2)
\le
-\frac{1}{2}\log\frac{\sigma^2}{\sigma_0^2}
-\frac{1}{2}\Big(\frac{\sigma_0^2}{\sigma^2}-1\Big),
\]
where we also used $\frac{1}{m}\tr(\xmat{M}(\xmat{\Psi}))\ge 1$ from \eqref{eq:trlog_pf}.
The scalar inequality $a-\log a-1\ge0$ for $a>0$ (with $a=\sigma^2/\sigma_0^2$) implies that the right-hand side is
non-positive, with equality if and only if $\sigma^2=\sigma_0^2$.
Hence \eqref{eq:ident_pf} holds.

If equality holds, then necessarily $\sigma^2=\sigma_0^2$ and equality must hold in \eqref{eq:trlog_pf} and in
\eqref{eq:Ert_pf}. Thus $\xmat{M}(\xmat{\Psi})=\xmat{I}_m$, hence $\xmat{\Psi}=\xmat{\Psi}_0$, and
$\E[\Delta_t(\vartheta)^\top\Delta_t(\vartheta)]=0$, hence $\Delta_t(\vartheta)=0$ almost surely.
With $\xmat{\Psi}=\xmat{\Psi}_0$, $\Delta_t(\vartheta)=0$ reduces to
\[
\xvec{x}_t(\xmat{B})
+
\sum_{\ell=1}^{L}
\xmat{A}_{\ell}(\xmat{\Pi}_{\ell})
\ddot{\xvec{Y}}_{t-\tau_\ell}
=
\xvec{x}_t(\xmat{B}_0)
+
\sum_{\ell=1}^{L}
\xmat{A}_{0,\ell}
\ddot{\xvec{Y}}_{t-\tau_\ell}
\quad\text{a.s.}
\]
By Assumption~\ref{ass:rank_id_cons}, $\E(z_tz_t^\top)\succ0$, and since the map
$(\vop(\xmat{B}),\vop(\xmat{\Pi}_1),\dots,\vop(\xmat{\Pi}_L)) \mapsto \xvec{x}_t(\xmat{B}) + \sum_{\ell=1}^{L} \xmat{A}_{\ell}(\xmat{\Pi}_{\ell}) \ddot{\xvec{Y}}_{t-\tau_\ell}$ is linear,
it follows that $\xmat{B}=\xmat{B}_0$ and $\xmat{\Pi}_{\ell}=\xmat{\Pi}_{0,\ell}$ for all $\ell=1,\dots,L$. Hence, $(\vartheta_0,\sigma_0^2)$ is uniquely identified.

\medskip

\noindent\textbf{Step 2: Uniform convergence.}
We show that
\[
\sup_{(\vartheta,\sigma^2)\in\Theta\times[\underline\sigma^2,\overline\sigma^2]}
\big|\ell_{nT}(\vartheta,\sigma^2)-Q(\vartheta,\sigma^2)\big|\xrightarrow{p}0.
\]
The term $-\frac12\log\sigma^2$ is continuous and bounded on $[\underline\sigma^2,\overline\sigma^2]$.
By Assumption~\ref{ass:stability_cons}, the map $\xmat{\Psi}\mapsto \frac{1}{m}\log|\xmat{S}(\xmat{\Psi})|$ is continuous and uniformly bounded on $\Theta$.

It remains to establish a uniform law of large numbers for
\[
G_{nT}(\vartheta)=\frac{1}{mT}\sum_{t=1}^T \xvec{r}_t(\vartheta)^\top \xvec{r}_t(\vartheta).
\]
The following argument follows standard uniform convergence results for M-estimators under stationarity and ergodicity; see, for example, \cite{newey1994large} or \cite{white1996estimation}. Under Assumptions~\ref{ass:err_exo_cons}, \ref{ass:stability_cons} and \ref{ass:asymptotics_cons}
(and, if the regressors are stochastic, Assumption~\ref{ass:X_stoch_cons}),
the process $\{(\ddot{\xvec{Y}}_t,\xmat{X}_t)\}$ is strictly stationary and ergodic with finite $(2+\delta)$ moments for some $\delta>0$.
Therefore, for each fixed $\vartheta\in\Theta$,
\[
G_{nT}(\vartheta)\xrightarrow{p}\E\!\left[\frac{1}{m}\xvec{r}_t(\vartheta)^\top \xvec{r}_t(\vartheta)\right].
\]

To obtain uniform convergence over compact $\Theta$, note that $\xvec{r}_t(\vartheta)$ is affine in \linebreak
$(\xmat{S}(\xmat{\Psi}),\xmat{B},\xmat{\Pi}_1,\dots,\xmat{\Pi}_L)$ and linear in $(\ddot{\xvec{Y}}_t,\ddot{\xvec{Y}}_{t-\tau_1}, \ddot{\xvec{Y}}_{t-\tau_L},\xmat{X}_t)$.
Compactness of $\Theta$ and the uniform bounds on $\xmat{S}(\xmat{\Psi})$ and $\xmat{S}(\xmat{\Psi})^{-1}$
(Assumption~\ref{ass:stability_cons}) imply that there exists $C<\infty$ such that, uniformly in $\vartheta_1,\vartheta_2\in\Theta$,
\[
\|\xvec{r}_t(\vartheta_1)-\xvec{r}_t(\vartheta_2)\|
\le
C\|\vartheta_1-\vartheta_2\|\big(\|\ddot{\xvec{Y}}_t\|+ \sum_{\ell=1}^{L}\|\ddot{\xvec{Y}}_{t-\tau_\ell}\| +\|\xmat{X}_t\|\big),
\]
and $\sup_{\vartheta\in\Theta}\|\xvec{r}_t(\vartheta)\|\le C(\|\ddot{\xvec{Y}}_t\|+ \sum_{\ell=1}^{L}\|\ddot{\xvec{Y}}_{t-\tau_\ell}\| +\|\xmat{X}_t\|+\|\xvec{u}_t\|)$.
Consequently,
\begin{eqnarray*}
|G_{nT}(\vartheta_1)-G_{nT}(\vartheta_2)| 
\le
\|\vartheta_1-\vartheta_2\|\cdot L_{nT},
\\
L_{nT} =\frac{C}{T}\sum_{t=1}^T\big(\|\ddot{\xvec{Y}}_t\|+ \sum_{\ell=1}^{L}\|\ddot{\xvec{Y}}_{t-\tau_\ell}\| +\|\xmat{X}_t\|+\|\xvec{u}_t\|\big)^2. 
\end{eqnarray*}
By ergodicity and finite moments, $L_{nT}=O_p(1)$.
Cover $\Theta$ by a finite $\varepsilon$-net $\{\vartheta^{(1)},\ldots,\vartheta^{(N_\varepsilon)}\}$.
Then
\[
\sup_{\vartheta\in\Theta}\big|G_{nT}(\vartheta)-\E[G_{nT}(\vartheta)]\big|
\le
\max_{k\le N_\varepsilon}\big|G_{nT}(\vartheta^{(k)})-\E[G_{nT}(\vartheta^{(k)})]\big|
+2\varepsilon L_{nT}.
\]
Pointwise convergence at the net points and $L_{nT}=O_p(1)$ imply
\[
\sup_{\vartheta\in\Theta}\big|G_{nT}(\vartheta)-\E[G_{nT}(\vartheta)]\big|\xrightarrow{p}0
\quad\text{as }T\to\infty,
\]
after letting $\varepsilon\downarrow 0$.
Combining with the bounded continuous terms yields the desired uniform convergence of $\ell_{nT}$ to $Q$.

\medskip
\noindent\textbf{Step 3: Consistency.}
By Step~1, $Q(\vartheta,\sigma^2)$ is uniquely maximised at $(\vartheta_0,\sigma_0^2)$.
By Step~2, $\ell_{nT}(\vartheta,\sigma^2)$ converges uniformly in probability to $Q(\vartheta,\sigma^2)$ on the compact
domain $\Theta\times[\underline\sigma^2,\overline\sigma^2]$.
Therefore, by a standard argmax theorem \citep[Theorem 2.1]{newey1994large},
\[
(\hat\vartheta_{nT},\hat\sigma^2_{nT})\xrightarrow{p}(\vartheta_0,\sigma_0^2),
\]
which completes the proof.
\end{proof}

\begin{proof}[Proof of Theorem~\ref{thm:AN}]
Let $\eta=(\vartheta^\top,\sigma^2)^\top$ denote the stacked parameter
vector and $\eta_0=(\vartheta_0^\top,\sigma_0^2)^\top$ the true parameter.
Consistency of $(\hat\vartheta_{nT},\hat\sigma^2_{nT})$ follows from
Theorem~\ref{th:consistency}, implying $\hat\eta_{nT}\xrightarrow{p}\eta_0$.

Recall that $\ell_{nT}(\eta)=\frac{1}{T}\sum_{t=1}^T\ell_t(\eta)$,
where $\ell_t(\eta)$ denotes the single-time contribution to the
criterion function. The first-order condition for the QML estimator yields
\[
\xvec{0}
=
\nabla_\eta \ell_{nT}(\hat\eta_{nT})
=
\nabla_\eta \ell_{nT}(\eta_0)
+
\left[\nabla^2_{\eta\eta}\ell_{nT}(\tilde\eta_{nT})\right]
(\hat\eta_{nT}-\eta_0),
\]
where $\tilde\eta_{nT}$ lies on the line segment between
$\hat\eta_{nT}$ and $\eta_0$.

Rearranging gives the linear expansion
\begin{equation}\label{eq:AN_linexp2}
\sqrt{T}(\hat\eta_{nT}-\eta_0)
=
-\left[\nabla^2_{\eta\eta}\ell_{nT}(\tilde\eta_{nT})\right]^{-1}
\cdot
\sqrt{T}\,\nabla_\eta\ell_{nT}(\eta_0).
\end{equation}

By Assumption~\ref{ass:AN_reg}(ii),
\[
\sqrt{T}\,\nabla_\eta\ell_{nT}(\eta_0)
=
\frac{1}{\sqrt{T}}\sum_{t=1}^T\nabla_\eta\ell_t(\eta_0)
\xrightarrow{d}N(\xvec{0},\mathcal{I}).
\]

By the structural properties of the likelihood and the regularity
conditions stated above, the Hessian of the sample criterion converges
locally uniformly in probability to its expectation. Hence,
\[
\sup_{\eta\in\mathcal{N}(\eta_0)}
\left\|
\nabla^2_{\eta\eta}\ell_{nT}(\eta)
-
\E[\nabla^2_{\eta\eta}\ell_t(\eta)]
\right\|
\xrightarrow{p}0.
\]
Since $\tilde\eta_{nT}\xrightarrow{p}\eta_0$,
it follows that
\[
\nabla^2_{\eta\eta}\ell_{nT}(\tilde\eta_{nT})
=
\E[\nabla^2_{\eta\eta}\ell_t(\tilde\eta_{nT})]
+o_p(1).
\]

By continuity of
$\eta\mapsto\E[\nabla^2_{\eta\eta}\ell_t(\eta)]$ at $\eta_0$,
\[
\E[\nabla^2_{\eta\eta}\ell_t(\tilde\eta_{nT})]
\xrightarrow{p}
\E[\nabla^2_{\eta\eta}\ell_t(\eta_0)]
=
-\mathcal{J}.
\]

Hence
\[
\nabla^2_{\eta\eta}\ell_{nT}(\tilde\eta_{nT})
\xrightarrow{p}-\mathcal{J}.
\]

By Assumption~\ref{ass:AN_reg}(i),
$\mathcal{J}$ is positive definite and therefore nonsingular, implying
\[
\left[\nabla^2_{\eta\eta}\ell_{nT}(\tilde\eta_{nT})\right]^{-1}
\xrightarrow{p}
-\mathcal{J}^{-1}.
\]

Applying Slutsky's theorem to \eqref{eq:AN_linexp2} yields
\[
\sqrt{T}(\hat\eta_{nT}-\eta_0)
\xrightarrow{d}
N\!\left(\xvec{0},\,\mathcal{J}^{-1}\mathcal{I}\mathcal{J}^{-1}\right).
\]

Under correct Gaussian specification,
the information identity implies $\mathcal{I}=\mathcal{J}$,
which yields the simplified covariance matrix $\mathcal{J}^{-1}$.
\end{proof}


\bibliographystyle{apalike} 

\end{document}